\documentclass[12pt]{article}
\usepackage{amssymb}

\begin{document}

\newcommand{\sect}[1]{\setcounter{equation}{0}\section{#1}}
\renewcommand{\theequation}{\thesection.\arabic{equation}}
\newcommand{\be}{\begin{equation}}
\newcommand{\ee}{\end{equation}}
\newcommand{\bea}{\begin{eqnarray}}
\newcommand{\eea}{\end{eqnarray}}
\newcommand{\nonu}{\nonumber\\}
\newcommand{\beano}{\begin{eqnarray*}}
\newcommand{\eeano}{\end{eqnarray*}}
\newcommand{\eps}{\epsilon}
\newcommand{\om}{\omega}
\newcommand{\vph}{\varphi}
\newcommand{\sig}{\sigma}
\newcommand{\CC}{\mbox{${\mathbb C}$}}
\newcommand{\RR}{\mbox{${\mathbb R}$}}
\newcommand{\QQ}{\mbox{${\mathbb Q}$}}
\newcommand{\ZZ}{\mbox{${\mathbb Z}$}}
\newcommand{\NN}{\mbox{${\mathbb N}$}}
\newcommand{\1}{\mbox{\hspace{.0em}1\hspace{-.24em}I}}
\newcommand{\II}{\mbox{${\mathbb I}$}}
\newcommand{\prt}{\partial}
\newcommand{\und}[1]{\underline{#1}}
\newcommand{\wh}[1]{\widehat{#1}}
\newcommand{\wt}[1]{\widetilde{#1}}
\newcommand{\mb}[1]{\ \mbox{\ #1\ }\ }
\newcommand{\half}{\frac{1}{2}}
\newcommand{\noin}{\not\!\in}
\newcommand{\rhotimes}{\mbox{\raisebox{-1.2ex}{$\stackrel{\displaystyle\otimes}
{\mbox{\scriptsize{$\rho$}}}$}}}
\newcommand{\bin}[2]{{\left( {#1 \atop #2} \right)}}
\newcommand{\A}{{\cal A}}
\newcommand{\B}{{\cal B}}
\newcommand{\C}{{\cal C}}
\newcommand{\F}{{\cal F}}
\newcommand{\R}{{\cal R}}
\newcommand{\T}{{\cal T}}
\newcommand{\W}{{\cal W}}
\newcommand{\cS}{{\cal S}}
\newcommand{\bS}{{\bf S}}
\newcommand{\cL}{{\cal L}}
\newcommand{\hlp}{{\RR}_+}
\newcommand{\hlm}{{\RR}_-}
\newcommand{\Hil}{{\cal H}}
\newcommand{\D}{{\cal D}}
\newcommand{\bg}{{\bf g}}
\newcommand{\alg}{\C_{\cS}}
\newcommand{\balg}{\B_{\cS}}
\newcommand{\frep}{\F_{\R,\T}(\C_\cS)}
\newcommand{\frepb}{\F_{\R,\T}(\C_\B)}
\newcommand{\rep}{\F(\C_\cS)}
\newcommand{\form}{\langle \, \cdot \, , \, \cdot \, \rangle }
\newcommand{\e}{{\rm e}}
\newcommand{\LL}{\mbox{${\mathbb L}$}}
\newcommand{\Rp}{{R^+_{\, \, \, \, }}}
\newcommand{\Rm}{{R^-_{\, \, \, \, }}}
\newcommand{\Rpm}{{R^\pm_{\, \, \, \, }}}
\newcommand{\Tp}{{T^+_{\, \, \, \, }}}
\newcommand{\Tm}{{T^-_{\, \, \, \, }}}
\newcommand{\Tpm}{{T^\pm_{\, \, \, \, }}}
\newcommand{\baral}{\bar{\alpha}}
\newcommand{\barbt}{\bar{\beta}}
\newcommand{\supp}{{\rm supp}\, }
\newcommand{\EE}{\mbox{${\mathbb E}$}}
\newcommand{\JJ}{\mbox{${\mathbb J}$}}
\newcommand{\MM}{\mbox{${\mathbb M}$}}
\newcommand{\ct}{{\cal T}}

\newtheorem{theo}{Theorem}[section]
\newtheorem{coro}[theo]{Corollary}
\newtheorem{prop}[theo]{Proposition}
\newtheorem{defi}[theo]{Definition}
\newtheorem{conj}[theo]{Conjecture}
\newtheorem{lem}[theo]{Lemma}
\newcommand{\prf}{\underline{\it Proof.}\ }
\newcommand{\finprf}{\null \hfill {\rule{5pt}{5pt}}\\[2.1ex]\indent}

\pagestyle{empty}
\rightline{March 2003}

\vfill

\begin{center}
{\Large\bf Reflection--Transmission Algebras}
\\[2.1em]

{\large
M. Mintchev$^{a}$\footnote{mintchev@df.unipi.it},
E. Ragoucy$^{b}$\footnote{ragoucy@lapp.in2p3.fr}
and P. Sorba$^{c}$\footnote{sorba@cern.ch and sorba@lapp.in2p3.fr, on leave of
      absence from LAPTH}}\\
\end{center}

\null

\noindent
{\it $^a$ INFN and Dipartimento di Fisica, Universit\'a di
       Pisa, Via Buonarroti 2, 56127 Pisa, Italy\\[2.1ex]
$^b$ LAPTH, 9, Chemin de Bellevue, BP 110, F-74941 Annecy-le-Vieux
       cedex, France\\[2.1ex]
   $^c$ TH Division, CERN, CH 1211 Geneva 23, Switzerland}
\vfill

\begin{abstract}

Inspired by factorized scattering from delta--type impurities in
(1+1)-dimen\-sional space-time, we propose and analyse a generalization
of the Zamolodchikov--Faddeev algebra. Distinguished elements of the
new algebra, called reflection and transmission generators, encode the
particle--impurity interactions. We describe in detail the underlying
algebraic structure. The relative Fock representations are
explicitly constructed and a general factorized scattering theory
is developed in this framework.

\end{abstract}

\vfill
\rightline{CERN-TH/2003-058}
\rightline{IFUP-TH 14/2003}
\rightline{LAPTH-972/03}
\rightline{\tt hep-th/0303187}
\newpage
\pagestyle{plain}
\setcounter{page}{1}

\sect{Introduction}

Much progress has been made in the last two decades in understanding
the physical
properties and the mathematical structure of integrable quantum
systems in 1+1 dimensions.
The idea of factorized scattering, which can be traced back to the
pioneering work of
Yang \cite{Yang:1967bm}, plays a central r\^ole in most of the
significant developments
in this field. It has been recognized later, that
the algebraic structure in the basis of factorized scattering theory is the
Zamolodchikov--Faddeev (ZF) algebra
\cite{Zamolodchikov:xm}--\cite{Faddeev:zy}. This algebra
represents a powerful tool for deriving not only $S$-matrix
amplitudes, but also form--factors
of local operators \cite{Karowski:1978vz, Smirnov:vz}.

Integrable models with boundaries \cite{Cherednik:vs}--\cite{Liguori:1998xr}
or defects \cite{Delfino:1994nr}--\cite{Mintchev:2002zd} have recently
also been the subject of intense
study. Since factorized scattering turns out to be fundamental in
this context as well,
the natural problem that arises is to find the counterpart of the ZF algebra,
which works
in the presence of reflecting and transmitting impurities. The main goal of
the present paper is to introduce such an algebra, called in what follows
{\sl reflection--transmission} (RT) algebra. Our strategy is to generalize the
approach to integrable systems on the half-line developed in
\cite{Liguori:1998xr}.
Besides the particle creation and annihilation
operators, the RT algebra involves also reflection and transmission
(defect) generators.
In the Fock representation, the latter acquire  non-vanishing vacuum
expectation values, defined in terms of
the observable reflection and transmission amplitudes of a single
particle interacting
with the defect. Together with the two-body bulk scattering matrix,
these amplitudes form the physical input. The structure of
the RT algebra is inspired by some exactly
solvable integrable models with $\delta$-type impurities.
Apart from providing a useful test for the general setup, these systems find
concrete applications \cite{Saleur:1998hq, Castro-Alvaredo:2002dj} in
conductance problems.

The paper is organized as follows. In the next section we focus on quantum
inverse scattering with $\delta$-impurities. The general concept of RT algebra
is introduced in section 3. After a brief description of the basic
features of these
algebras, we construct the relative Fock representations. Section 4
is devoted to
scattering with impurities. We first establish the unitarity and factorization
constraints. Using the Fock representation of a suitable RT algebra, we define
afterwards the ``in"-coming and ``out"-going states and construct
the total scattering operator. The last
section contains our conclusions. We discuss there some universal features
of the RT algebras and their relevance to inverse scattering and
other related topics.

\sect {Origin of reflection--transmission algebras}

It is instructive to start the discussion with two examples, showing how the
RT algebras emerge from the study of quantum impurity problems.
We begin with the $n$-particle Hamiltonian
\be
H^{(n)} = \sum_{i=1}^n - \frac{1}{2} \prt_{x_i}^2 + \eta\, \delta (x_i) \, ,
\qquad \eta \in \RR\, ,
\label{h1}
\ee
which describes a system of $n$ non-relativistic bosonic particles (of
unit mass) on $\RR$, which
interact with a $\delta$-type impurity localized in the origin, but
not among themselves.
This model is well-known to be exactly solvable. It is sufficient to
investigate
the spectral problem associated with the one-particle Hamiltonian $H^{(1)}$,
defined on a suitable (see e.g. \cite{A}) domain $\D_\eta \subset
L^2(\RR , dx)$
of continuous functions on $\RR$, which are twice differentiable in
$\RR \setminus \{0\}$ and
satisfy
\be
\lim_{x\downarrow 0}\left [(\prt_x \psi )(x) - (\prt_x \psi )(-x) \right ] =
2\eta\, \psi (0) \, .
\label{bc}
\ee
$H^{(1)}$ is self-adjoint on $\D_\eta$.
A set of orthogonal (generalized) eigenstates
$\{\psi_k^\pm(x)\, :\, k\in \RR \}$, verifying (\ref{bc}), is
\be
\psi_k^\pm(x) = \theta(\mp k) \left \{\theta(\mp x) T(\mp k) \e^{ikx} +
\theta(\pm x)\left [\e^{ikx} + R(\mp k)\e^{-ikx}\right ] \right \}\, ,
\label{basis}
\ee
where $\theta$ denotes the standard Heaviside function and
\be
T(k) = \frac{k}{k + i \eta}\, , \qquad
R(k) = \frac{-i\eta}{k + i \eta}\, .
\label{TR1}
\ee
The family $\{{\overline \psi}_{-k}^{\, \pm}(x)\, :\, k\in \RR \}$,
where the bar stands for complex conjugation, is also
orthonormal. The systems $\{\psi_k^\pm(x)\, :\, k\in \RR \}$ and
$\{{\overline \psi}_{-k}^{\, \pm}(x)\, :\, k\in \RR \}$ represent
physically scattering
states and, for $\eta \geq 0$, are separately complete in $L^2(\RR , dx)$.
When $\eta <0$, there is in addition one bound state
\be
\psi_{\rm b}(x) = \theta (-\eta)\sqrt {|\eta |} \left [\theta (x) \e^{\eta x} +
\theta (-x) \e^{-\eta x}\right ]\, ,
\label{bs}
\ee
which is orthogonal to $\{\psi_k^\pm(x)\, :\, k\in \RR \}$ and
$\{{\overline \psi}_{-k}^{\, \pm}(x)\, :\, k\in \RR \}$.
The energy of $\psi_{\rm b}$ is $E=-\eta^2/2$.

Particle collision and production processes are absent from this simple model.
Nevertheless, the reflection and transmission from the impurity give rise to a
non-trivial scattering operator, which preserves the particle number and
can be described as follows. Using the {\sl weak} limits
\be
\lim_{x\to \pm \infty} \frac {\e^{\pm ikx }}{k+i\epsilon} = 0 \, ,
\qquad \epsilon > 0\, ,
\label{wlim}
\ee
one can verify that
\be
\lim_{x\to \pm \infty} [\psi_k^\pm (x) - \e^{ikx}] = 0\, .
\label{asymp}
\ee
Therefore, one can interpret $\psi_k^\pm (x)$
asymptotically as incoming waves, travelling in $\RR_\pm$
with momentum $k\not= 0$ towards the impurity. Accordingly, we take the vectors
\be
|k\rangle^{\rm in} = \psi_k^+ (x) + \psi_k^- (x) \, ,
\label{opin}
\ee
to be the basis of one-particle ``in"  states. Analogous considerations
lead us to choose the following basis of one-particle ``out"  states:
\be
|k\rangle^{\rm out}  = {\overline \psi}_{-k}^{\, +} (x) +
{\overline \psi}_{-k}^{\, -} (x) \, .
\label{opout}
\ee
The one-particle scattering operator is defined at this point by
\be
\bS^{(1)}\,
|k\rangle^{\rm out} = |k\rangle^{\rm in} \, .
\label{opS}
\ee
By construction, $\bS^{(1)}$ is a unitary operator on
$L^2(\RR , dx)$ for $\eta \geq 0$. In the range $\eta < 0$,
$\bS^{(1)}$ is defined and
unitary on the orthogonal complement to the bound state (\ref{bs}) in
$L^2(\RR , dx)$.
The one-particle transition amplitude reads
\bea
&{}^{\rm out}\langle p|k\rangle^{\rm in} =
{}^{\rm out}\langle p|\bS^{(1)}|k\rangle^{\rm out} = \nonumber \\
&[\theta (p) T(p) + \theta (-p) T(-p)]2\pi \delta (p-k) + \nonumber \\
&[\theta (p) R(p) + \theta (-p) R(-p)]2\pi \delta (p+k) \, ,
\label{opta}
\eea
which clarifies the physical meaning of $T$ and $R$, given by eq. (\ref{TR1}).
They represent the transmission and reflection amplitudes and admit a
meromorphic continuation
in $k$ to the whole complex plane $\CC$. The pole $k=-i\eta$ confirms
the presence of
the bound state $\psi_{\rm b}$ for $\eta<0$ and indicates the existence of a
resonance state for $\eta >0$.
 
The $n$-particle amplitude, with initial and final configurations satisfying
$k_1<...<k_n$ and $p_1>...>p_n$ respectively, can be expressed in
terms of (\ref{opta})
as follows:
\be
{}^{\rm out}\langle p_1,...,p_n|k_1,...,k_n\rangle^{\rm in} =
{}^{\rm out}\langle p_1,...,p_n|\bS^{(n)}|k_1,...,k_n\rangle^{\rm out} =
\prod_{i=1}^n {}^{\rm out}\langle p_i|k_i\rangle^{\rm in} \, .
\label{ampl1}
\ee
Equation (\ref{ampl1}) concludes our brief summary of the standard
and well-known
analytic treatment of the integrable system defined by (\ref{h1}).

A natural question one may ask at this point concerns the existence of an
algebraic framework for dealing with the above system for $\eta\not=0$,
similar to the familiar canonical commutation approach, which works in the case
$\eta = 0$. The answer to this question turns out to be
affirmative and we now turn to the description of the relevant
algebraic structure. Following
our previous work \cite{Mintchev:2002zd}, we introduce the associative
algebra $\C_B$  with identity element $\bf 1$, generated by
$\{a^{\ast \xi} (k),\, a_\xi (k)\, :\, \xi=\pm,\, k\in \RR \}$
obeying the bosonic-type commutation relations:
\bea
&a_{\xi_1}(k_1)\, a_{\xi_2}(k_2) -  a_{\xi_2}(k_2)\, a_{\xi_1}(k_1) = 0\, ,
\label{ccr1} \\
&a^{\ast \xi_1}(k_1)\, a^{\ast \xi_2}(k_2) - a^{\ast \xi_2}(k_2)\,
a^{\ast \xi_1}(k_1) = 0\, ,
\label{ccr2} \\
&a_{\xi_1}(k_1)\, a^{\ast \xi_2}(k_2) - a^{\ast \xi_2}(k_2)\,
a_{\xi_1}(k_1) = \nonumber \\
&\left [\delta_{\xi_1}^{\xi_2} + \T_{\xi_1}^{\xi_2}(k_1)\right ] 2\pi
\delta(k_1-k_2)\, {\bf 1} +
\R_{\xi_1}^{\xi_2}(k_1) 2\pi \delta(k_1+k_2)\, {\bf 1}\, ,
\label{ccr3}
\eea
where
\be
\T(k) =\left(\begin{array}{cc} 0&T(k)\\ {\overline
T}(k)&0\end{array}\right)\, ,
\qquad
\R(k) =\left(\begin{array}{cc} R(k)&0\\0&{\overline R}(k)\end{array}\right)\, .
\label{TR2}
\ee
The right-hand side of eq. (\ref{ccr3}) captures the presence of the impurity.
The term proportional to $\delta(k_1+k_2)$ reflects in particular the
breaking of translation and Galilean invariance due to the impurity.
We shall see
in the next section that $\C_B$ is a particular RT algebra. For
the moment we focus on the Fock
representation $\frepb$ of $\C_B$, referring for the explicit
construction to sect. 3.2.
An essential feature of $\frepb$ is that the operators $\{a^{\ast
\xi} (k),\, a_\xi (k)\}$ in
this representation satisfy
\bea
a_\xi(k) &=& \T_\xi^\eta (k) a_\eta (k) + \R_\xi^\eta (k) a_\eta (-k) \, ,
\label{c1} \\
a^{\ast \xi}(k) &=& a^{\ast \eta}(k) \T_\eta^\xi (k) +
a^{\ast \eta}(-k) \R_\eta^\xi (-k) \, .
\label{c2}
\eea
Hereafter the summation over repeated upper and lower indices is understood.
The relations (\ref{c1}), (\ref{c2}) originate from
the reflection--transmission automorphism characterizing any
RT algebra and established in sect. 3.
In the physical context these relations encode the interaction with
the impurity.

The vacuum state $\Omega \in \frepb$ obeys as usual $a_\xi (k)\Omega = 0$. We
denote by $( \cdot\, ,\, \cdot )$ the scalar product in $\frepb$
and consider the vacuum expectation value
\be
(a^{\ast \eta_1}(p_1)...a^{\ast \eta_n}(p_n)\Omega
\, , \, a^{\ast \xi_1}(k_1)...a^{\ast \xi_n}(k_n)\Omega ) \, ,
\label{ampl2}
\ee
with
\be
k_1<\cdots <k_n\, ,\quad \xi_i = -\epsilon (k_i)\, ,
\qquad p_1>\cdots >p_n\, ,\quad \eta_i = \epsilon (p_i) \, ,
\label{cond1}
\ee
$\epsilon$ being the sign function. By means of eqs. (\ref{ccr1})--(\ref{ccr3})
it is easily verified that (\ref{ampl2}) precisely reproduce
the amplitudes (\ref{ampl1}) for any $n$. Therefore, $\C_B$ provides a purely
algebraic framework for constructing the scattering operator. The
formalism actually applies
to any observable of the system, introducing in addition to $\C_B$
the creation and annihilation operators $\{b^\ast\, ,\, b\}$ for
the bound state (\ref{bs}), which commute with $\{a^{\ast
\xi} (k),\, a_\xi (k)\}$  and satisfy
\be
[b\, ,\, b] = [b^\ast\, ,\, b^\ast] = 0\, , \qquad [b\, ,\, b^\ast ] = 1 \, .
\ee
{}For the Hamiltonian one finds, for instance,
\be
H = \frac{1}{2} \int_{-\infty}^{+\infty} \frac{dk}{2\pi}k^2 a^{\ast
\xi}(k) a_\xi (k) -
\theta(-\eta)\frac{\eta^2}{2}b^\ast b \, .
\label{h2}
\ee
The restriction of $H$ to the $n$-particle subspace of the total
Hilbert space (including
the bound state) is the algebraic counterpart of the Hamiltonian
(\ref{h1}) we started with.

At this stage we have enough background to turn to quantum field
theory with $\delta $-type impurities \cite{Delfino:1994nr,
Castro-Alvaredo:2002dj}.
Our goal will be to demonstrate that the algebra $\C_B$ can
be successfully applied there as well. As an example we consider the model
\bea
&S[\varphi] = \frac{1}{2} \int_{-\infty}^{+\infty} dt
\int_{-\infty}^{+\infty}dx
[(\prt_t\varphi )^2(t,x) - (\prt_x\varphi )^2(t,x) \nonumber \\
&- m^2 \varphi^2 (t,x) - 2\eta \delta (x) \varphi^2 (t,x)] \, ,
\label{action}
\eea
with $m\geq 0$ and $\eta \in \RR$. The action (\ref{action}) defines a standard
external field problem with $\delta$-potential. The corresponding
equation of motion is
\be
[\prt_t^2 - \prt_x^2 + m^2 + 2\eta \delta (x)] \varphi (t,x) = 0\, ,
\label{eqm}
\ee
and our problem will now be to quantize (\ref{eqm}) with the standard
initial conditions:
\be
[\varphi (0,x_1)\, ,\, \varphi (0,x_2)] = 0\, , \qquad
[(\prt_t\varphi )(0,x_1)\, ,\, \varphi (0,x_2)] = -i\delta (x_1-x_2) \, .
\label{initial}
\ee
The solution of this problem requires the study of the operator
\be
K \equiv -\prt_x^2 + m^2 +2\eta \delta (x) \, .
\label{K}
\ee
We already know that $K$ is self-adjoint on $\D_\eta$. In order to
avoid imaginary energies,
we demand $K$ to be non-negative, which implies
\be
-m \leq \eta \, .
\label{stab}
\ee
Now, the solution of eqs. (\ref{eqm}) and (\ref{initial})
is unique and can be expressed in terms of the generators
$\{a^{\ast \xi} (k),\, a_\xi (k)\}$ and $\{b^\ast\, ,\, b\}$.
One finds
\be
\varphi (t,x) = \varphi_+ (t,x) + \varphi_- (t,x) + \varphi_{\rm b} (t,x) \, ,
\label{f1}
\ee
where
\be
\varphi_\pm (t,x) = \int_{-\infty}^{+\infty} \frac{dk}{2\pi \sqrt
{2\omega (k)}}
\left[a^{\ast \pm}(k) {\overline \psi}_k^{\, \pm} (x)\e^{i\omega (k)t} +
a_\pm (k) \psi_k^\pm (x)\e^{-i\omega (k)t}\right ] \, ,
\label{f2}
\ee
\be
\varphi_{\rm b} (t,x) = \frac{1}{\sqrt {2\omega (i\eta)}}
\left [b^\ast\, \e^{it\omega (i\eta)} +
b\, \e^{-it\omega (i\eta)}\right ] \psi_{\rm b} (x)\, ,
\label{f3}
\ee
with $\omega (k) = \sqrt {k^2 + m^2}$.

Using eqs. (\ref{ccr3}) and (\ref{f1})--(\ref{f3}), one easily derives
the two-point vacuum expectation value
\bea
&&w^{(2)}(t_1,x_1,t_2,x_n) =
(\varphi (t_1,x_1)\Omega\, ,\, \varphi (t_2,x_2)\Omega ) = \nonumber \\
&&\int_{-\infty}^{+\infty} \frac{dk}{4\pi\omega (k)} \e^{-i\omega (k)t_{12}}
\Big\{\theta (x_1)\theta (-x_2) T(k)\e^{ikx_{12}} +
\theta (-x_1)\theta (x_2) {\overline T}(k)\e^{ikx_{12}} + \nonumber
\\[.21ex]
&&\theta (x_1)\theta (x_2)
\left [\e^{ikx_{12}} + R(k)\e^{ik{\widetilde x}_{12}}\right ] +
\theta (-x_1)\theta (-x_2)
\left [\e^{ikx_{12}} + {\overline R}(k)\e^{ik{\widetilde
x}_{12}}\right] \Big\} + \nonumber \\[.21ex]
&&\frac{1}{2\omega (i\eta)}\e^{-it_{12}\omega (i\eta)} \psi_{\rm
b}(x_1)  \psi_{\rm b}(x_2) \, ,
\label{w2}
\eea
where $t_{12} = t_1-t_2$, $x_{12} = x_1-x_2$ and ${\widetilde x}_{12}
= x_1+x_2$.
The last term in (\ref{w2}) represents the contribution of the bound
state and vanishes for
$\eta \geq 0$. The field $\varphi$ has a relativistic dispersion relation
$\omega (k)^2 = k^2 + m^2$, but nevertheless Lorentz invariance is
manifestly broken in (\ref{w2}).

The function (\ref{w2}) fully determines the theory. In fact $w^{(2n+1)} = 0$,
whereas $w^{(2n)}$ can be derived from $w^{(2)}$ by means of the
well-known recursion relation
\bea
&w^{(2n)}(t_1,x_1,...,t_{2n},x_{2n}) = \nonumber \\
&\sum_{i=1}^{2n-1} w^{(2)}(t_i,x_i,t_{2n},x_{2n})\,
w^{(2n-2)}(t_1,x_1,...,{\widehat t}_i,{\widehat
x}_i,...,t_{2n-1},x_{2n-1}) \, , \qquad
\label{w2n}
\eea
the hat indicating that the corresponding argument must be omitted.

Having at our disposal all
correlation functions, we can derive the scattering operator of the theory.
The particles of our model do not interact directly, but interact with the
external $\delta$-function field, modelling the impurity. We will now show that
the associated scattering matrix is fully determined by
the algebra $\C_B$. Equation (\ref{f2}) therefore represents a true
quantum inverse scattering
transform, allowing a reconstruction of the fields $\varphi_\pm$ from
$\{a^{\ast \xi} (k),\, a_\xi (k)\}$.
Let us concentrate first on the case $\eta \geq 0$, commenting at the end
on the range
$-m\leq \eta <0$. In developing the scattering theory one can use the
Haag--Ruelle approach \cite{RSIII}
with some minor modifications \cite{Gattobigio:1998si}, which reflect
the absence of translation invariance.
The novel feature, with respect to the quantum mechanical example
discussed above, is
that in quantum field theory we need smearing with special
wave-packets for the free Klein--Gordon
equation, which keep trace of the position $x=0$ of the impurity.
Such wave-packets
can be introduced as follows. Let $\D(\RR)$ be the space of smooth
test functions
with compact support. Then
\be
f^t (x) = \int_{-\infty}^{+\infty} \frac{dk}{2\pi \sqrt {2\omega
(k)}} f(k) \e^{ikx-i\omega (k)t} \, ,
\qquad f\in \D(\RR)\, ,
\label{KG}
\ee
is a smooth solution of the Klein--Gordon equation of mass $m$. We
will say that
$f_1\in \D(\RR)$ precedes $f_2\in \D(\RR)$ and write $f_1\prec f_2$
if and only if
$\supp (f_1) \cap \supp (f_2) = \emptyset $ and $k_1 < k_2$ for all
$k_1\in \supp (f_1)$ and all\break $k_2\in \supp (f_2)$. We now introduce
   the two sets
$\{g_i(k) \in \D(\RR)\, :\, i=1,...,m\}$ and $\{h_j(k) \in \D(\RR)\,
:\, j=1,...,n\}$,
which satisfy the non-overlapping conditions
\bea
g_1 \prec \cdots \prec g_m \, , &\qquad
h_n \prec \cdots \prec h_1 \, , \nonumber \\
0\not\in \supp g_i \, , \, \, \,  &\qquad 0\not\in \supp h_j \, .
\label{nonoverlap}
\eea
Setting now
\be
\xi_i =
\left\{ \begin{array}{cc}
+ \, , & \quad \mbox{$\supp g_i \subset \RR_-$}\, ,
\\ -\, , & \quad \mbox{$\supp g_i \subset \RR_+$}\, ,
\end{array} \right. \qquad
\eta_j =
\left\{ \begin{array}{cc}
+ \, , & \quad \mbox{$\supp h_j \subset \RR_+$}\, ,
\\ -\, , & \quad \mbox{$\supp h_j \subset \RR_-$}\, ,
\end{array} \right.
\label{xieta}
\ee
we define
\be
g^t_{\xi_i}(x) = \theta (\xi_i x)\, g_i^{\, t} (x) \, , \qquad
h^t_{\eta_j}(x) = \theta (\eta_j x)\, h_j^{\, t} (x) \, .
\label{gxiheta}
\ee
By construction, $g^t_{\xi_i}(x)$ represent wave-packets in
$\RR_{\xi_i}$ which move towards the impurity in $x=0$. On the other hand,
$h^t_{\eta_j}(x)$ are wave-packets in $\RR_{\eta_j}$, which travel
away from the impurity in the direction $x\to \eta_j \infty$. Therefore
one expects that the smeared operators
\be
\varphi (t, g_{\xi_i}^t) = i \int_{-\infty}^{+\infty} dx\,
[(\prt_t g_{\xi_i}^t) (x)\, \varphi (t,x) -
g_{\xi_i}^t(x)\, (\prt_t \varphi) (t,x) ] \, ,
\label{phiin}
\ee
\be
\varphi (t, h_{\eta_j}^t) = i \int_{-\infty}^{+\infty} dx\,
[(\prt_t h_{\eta_j}^t) (x)\, \varphi (t,x) -
h_{\eta_j}^t(x)\, (\prt_t \varphi) (t,x)]
\label{phiout}
\ee
generate asymptotic ``in" and ``out"  states respectively. This is
indeed the case because
of the existence of the following strong limits in the Fock space $\frepb$:
\bea
\lim_{t\to -\infty} \varphi (t, g_{\xi_1}^t) \cdots \varphi (t,
g_{\xi_m}^t) \Omega =
a^{\ast \xi_1}(g_1) \cdots a^{\ast \xi_m}(g_m) \Omega \equiv
|g_1,...,g_m \rangle^{\rm in} , \label{limin} \\
\lim_{t\to +\infty} \varphi (t, h_{\eta_1}^t) \cdots \varphi (t,
h_{\eta_n}^t) \Omega =
a^{\ast \eta_1}(h_1) \cdots a^{\ast \eta_n}(h_n) \Omega \equiv
|h_1,...,h_n \rangle^{\rm out} ,
\label{limout}
\eea
where
\be
a^{\ast \zeta}(f) = \int_{-\infty}^{+\infty} \frac{dk}{2\pi} f(k)
a^{\ast \zeta}(k) \, .
\ee
Let us sketch the proof of (\ref{limin}), for example. Using the
non-overlapping conditions
(\ref{nonoverlap}), the commutation relations
(\ref{ccr1})--(\ref{ccr3}) and the
constraints (\ref{c1}), (\ref{c2}), one first derives the identity
\be
\varphi (t, g_{\xi_1}^{t}) \cdots \varphi (t, g_{\xi_m}^t) \Omega =
a^{\ast \xi_1}({\widetilde g}_{\xi_1}^t) \cdots a^{\ast
\xi_m}({\widetilde g}_{\xi_m}^t) \Omega \, ,
\label{i1}
\ee
where
\be
{\widetilde g}_{\xi_i}^t (p) = \int_{-\infty}^{\infty} dx\, \theta (\xi_i x)
\int_{-\infty}^{\infty} \frac{dk}{2\pi}
\, \frac{\omega (k) + \omega (p)}{2\sqrt {\omega (k) \omega (p)}}\, g_i(k)
\, \e^{i[\omega (p)-\omega (k)]t - ix(p-k)} \, .
\label{gtilde}
\ee
Therefore, eq. (\ref{limin}) is equivalent to
\be
\lim_{t\to -\infty}
\parallel a^{\ast \xi_1}({\widetilde g}_{\xi_1}^t) \cdots a^{\ast
\xi_m}({\widetilde g}_{\xi_m}^t) \Omega
- a^{\ast \xi_1}(g_1) \cdots a^{\ast \xi_m}(g_m) \Omega \parallel = 0 \, ,
\label{limin1}
\ee
$\parallel \cdot \parallel$ being the $L^2$-norm. For proving
(\ref{limin1}) one uses the continuity
of $a^{\ast \zeta}(f)$ in $f$ (see the estimate (\ref{est1}) below) and
\be
\lim_{t\to -\infty} \parallel {\widetilde g}_{\xi_i}^t - g_i\parallel = 0 \, ,
\qquad \forall \, \,  i = 1,...,m \, .
\label{L2lim}
\ee
{}Finally, (\ref{L2lim}) is a consequence of the weak limits (\ref{wlim}),
   $x$ being replaced by $t$. This concludes the argument.

Summarizing, the finite linear combinations of the vectors
\bea
&\{|g_1,...,g_m \rangle^{\rm in}\, : g_1 \prec \cdots \prec g_m \,
,\, m=1,2,... \}\, , \\
&\{|h_1,...,h_n \rangle^{\rm out}\, :\, h_n \prec \cdots \prec h_1 \,
,\, n=1,2,...\}
\eea
generate, after completion with respect to the scalar product in the
Fock space $\frepb$,
the asymptotic spaces $\Hil^{\rm in}$ and $\Hil^{\rm out}$. It turns out that
asymptotic completeness $\Hil^{\rm out} = \frepb = \Hil^{\rm in}$
holds for $\eta \geq 0$.
The transition amplitudes read
\be
{}^{\rm out}\langle h_1,...,h_n |g_1,...,g_m \rangle^{\rm in} = \delta_{mn}\,
\prod_{i=1}^n \int_{-\infty}^{+\infty} \frac{dp_i}{2\pi} \frac{dk_i}{2\pi}
{\overline h}_i(p_i)\,{}^{\rm out}\langle p_i|k_i\rangle^{\rm in}\, 
g_i(k_i)\, ,
\ee
where ${}^{\rm out}\langle p|k\rangle^{\rm in}$ is given by (\ref{opta}).
These results can be generalized to the case $-m< \eta <0$, except for
the property of asymptotic completeness, which is violated by the bound
state present in this range.

It is worth mentioning that the above framework applies with
straightforward modifications to fermionic systems
as well. The relevant algebra $\C_F$ is obtained in this case simply
by replacing
the commutators in the right-hand sides of eqs. (\ref{ccr1})--(\ref{ccr3}) with
anticommutators.

It is evident from the above considerations that $\C_B$ is a universal
and powerful tool for
handling $\delta$-type impurities, both in quantum mechanics and
quantum field theory.
One can view $\C_B$ ($\C_F$) as a central extension of the algebra of canonical
commutation (anticommutation) relations. A direct generalization,
emerging at this point,
is to substitute $\T(k) {\bf 1}$ and $\R(k) {\bf 1}$ in eq.
(\ref{ccr3}) with new generators
$t(k)$ and $r(k)$, which are no longer central elements. Moreover,
in the spirit of the ZF algebra, it is possible to
replace the bosonic (fermionic) exchange factor between $\{a^{\ast
\xi} (k),\, a_\xi (k)\}$
with a more general one. In this way one naturally arrives at the
general concept
of RT algebra, described in the next section.

\sect {Reflection--transmission algebras}

\subsection {Definition and general properties}

Inspired by the above treatment of $\delta$-impurities, we introduce
an associative algebra with identity element $\bf 1$ and two
types of generators, $\{a_\alpha (\chi ),\, a^{\ast \alpha } (\chi )\}$ and
$\{r_\alpha^\beta (\chi), \, t_\alpha^\beta (\chi)\}$,
called bulk and defect (reflection and transmission)
generators, respectively. We refer to $\chi \in \RR$ as a
spectral parameter. In the context of inverse scattering $\chi $
parametrizes the particle
dispersion relation (see eqs. (\ref{gen}) below). To be able to deal
with systems with
internal degrees of freedom, we adopt double indices $\alpha = (\xi ,
i)$. The component
$\xi = \pm$ indicates the half-line $\RR_\pm$ where the particle is
created or annihilated,
whereas $i=1,...,N$ parametrizes the internal (``isotopic") degrees
of freedom. The generators
$\{a_\alpha (\chi ),\, a^{\ast \alpha } (\chi ),\, r_\alpha^\beta
(\chi), \, t_\alpha^\beta (\chi)\}$
are subject to the following constraints:
\begin{itemize}
\item {} bulk exchange relations
\bea
a_{\alpha_1}(\chi_1) \, \, a_{\alpha_2 }(\chi_2) \, \; - \; \,
        \cS_{\alpha_2 \alpha_1 }^{\beta_1 \beta_2 }
        (\chi_2 , \chi_1)\,\, a_{\beta_2 }(\chi_2)\, a_{\beta_1 
}(\chi_1) & = & 0
        \, , \qquad \quad \label{aa}\\
a^{\ast \alpha_1} (\chi_1)\, a^{\ast \alpha_2 } (\chi_2) -
        a^{\ast \beta_2 } (\chi_2)\, a^{\ast \beta_1 } (\chi_1)\,
        \cS_{\beta_2 \beta_1 }^{\alpha_1 \alpha_2 }(\chi_2 , \chi_1) & = & 0
        \, , \qquad \quad \label{a*a*} \\
a_{\alpha_1 }(\chi_1)\, a^{\ast \alpha_2 } (\chi_2) \; - \;
        a^{\ast \beta_2 }(\chi_2)\,
        \cS_{\alpha_1 \beta_2 }^{\alpha_2 \beta_1 }(\chi_1 , \chi_2)\,
        a_{\beta_1 }(\chi_1) & = &  \nonumber\\
        2\pi\, \delta (\chi_1 - \chi_2)\,
        \left [\delta_{\alpha_1 }^{\alpha_2 }\, {\bf 1} +
t_{\alpha_1}^{\beta_2}(\chi_1)\right] +
        2\pi\, \delta (\chi_1 + \chi_2)\,  r_{\alpha_1 }^{\alpha_2 }(\chi_1)
        \, ;
\label{aa*}
\eea
\item {} defect exchange relations
\bea
\cS_{\alpha_1 \alpha_2}^{\gamma_2 \gamma_1}(\chi_1 , \chi_2)\,
r_{\gamma_1}^{\delta_1}(\chi_1)\,
\cS_{\gamma_2 \delta_1}^{\beta_1 \delta_2}(\chi_2 , -\chi_1)\,
r_{\delta_2}^{\beta_2}(\chi_2) = \nonumber\\
r_{\alpha_2}^{\gamma_2}(\chi_2)\,
\cS_{\alpha_1\gamma_2}^{\delta_2 \delta_1}(\chi_1 , -\chi_2)\,
r^{\gamma_1}_{\delta_1}(\chi_1)\,
\cS_{\delta_2\gamma_1}^{\beta_1 \beta_2}(-\chi_2 , -\chi_1)
\, ; \label{rr}
\eea
\bea
\cS_{\alpha_1 \alpha_2}^{\gamma_2 \gamma_1}(\chi_1 , \chi_2)\,
t_{\gamma_1}^{\delta_1}(\chi_1)\,
\cS_{\gamma_2 \delta_1}^{\beta_1 \delta_2}(\chi_2 , \chi_1)\,
t_{\delta_2}^{\beta_2}(\chi_2) = \nonumber\\
t_{\alpha_2}^{\gamma_2}(\chi_2)\,
\cS_{\alpha_1\gamma_2}^{\delta_2 \delta_1}(\chi_1 , \chi_2)\,
t^{\gamma_1}_{\delta_1}(\chi_1)\,
\cS_{\delta_2\gamma_1}^{\beta_1 \beta_2}(\chi_2 , \chi_1)
\, ; \label{tt}
\eea
\bea
\cS_{\alpha_1 \alpha_2}^{\gamma_2 \gamma_1}(\chi_1 , \chi_2)\,
t_{\gamma_1}^{\delta_1}(\chi_1)\,
\cS_{\gamma_2 \delta_1}^{\beta_1 \delta_2}(\chi_2 , \chi_1)\,
r_{\delta_2}^{\beta_2}(\chi_2) = \nonumber\\
r_{\alpha_2}^{\gamma_2}(\chi_2)\,
\cS_{\alpha_1\gamma_2}^{\delta_2 \delta_1}(\chi_1 , -\chi_2)\,
t^{\gamma_1}_{\delta_1}(\chi_1)\,
\cS_{\delta_2\gamma_1}^{\beta_1 \beta_2}(-\chi_2 , \chi_1)
\, ; \label{tr}
\eea
\item {} mixed exchange relations
\be
a_{\alpha_1}(\chi_1)\, r_{\alpha_2}^{\beta_2}(\chi_2) =
\cS_{\alpha_2 \alpha_1}^{\gamma_1\gamma_2}(\chi_2 , \chi_1)\,
r_{\gamma_2}^{\delta_2}(\chi_2)\,
\cS_{\gamma_1 \delta_2}^{\beta_2 \delta_1}(\chi_1 , -\chi_2)\,
a_{\delta_1}(\chi_1)
\, , \qquad \label{ar}
\ee
\be
r_{\alpha_1}^{\beta_1}(\chi_1)\, a^{\ast \alpha_2}(\chi_2) =
a^{\ast \delta_2}(\chi_2)\,
\cS_{\alpha_1 \delta_2}^{\gamma_2 \delta_1}(\chi_1 , \chi_2)\,
r_{\delta_1}^{\gamma_1}(\chi_1)\,
\cS_{\gamma_2\gamma_1}^{\beta_1 \alpha_2}(\chi_2 , -\chi_1)\,
\, , \qquad \label{ra*}
\ee
\be
a_{\alpha_1}(\chi_1)\, t_{\alpha_2}^{\beta_2}(\chi_2) =
\cS_{\alpha_2 \alpha_1}^{\gamma_1\gamma_2}(\chi_2 , \chi_1)\,
t_{\gamma_2}^{\delta_2}(\chi_2)\,
\cS_{\gamma_1 \delta_2}^{\beta_2 \delta_1}(\chi_1 , \chi_2)\,
a_{\delta_1}(\chi_1)
\, , \qquad \label{at}
\ee
\be
t_{\alpha_1}^{\beta_1}(\chi_1)\, a^{\ast \alpha_2}(\chi_2) =
a^{\ast \delta_2}(\chi_2)\,
\cS_{\alpha_1 \delta_2}^{\gamma_2 \delta_1}(\chi_1 , \chi_2)\,
t_{\delta_1}^{\gamma_1}(\chi_1)\,
\cS_{\gamma_2\gamma_1}^{\beta_1 \alpha_2}(\chi_2 , \chi_1)\,
\, . \qquad \label{ta*}
\ee
\end{itemize}
The exchange factor $\cS$ is required to satisfy some compatibility
conditions, which read as follows, in conventional tensor notation:
\be
\cS_{12} (\chi_1,\chi_2)\, \cS_{12}(\chi_2,\chi_1) = \II\otimes \II \, ,
\label{unit1}
\ee
\be
\cS_{12}(\chi_1,\chi_2) \cS_{23}(\chi_1,\chi_3) \cS_{12} (\chi_2,\chi_3)
= \cS_{23}(\chi_2,\chi_3) \cS_{12}(\chi_1,\chi_3) \cS_{23}(\chi_1,\chi_2)  \, .
\label{qyb}
\ee
Equation (\ref{unit1}) is know as the ``unitarity" condition, whereas
(\ref{qyb}) is the celebrated quantum Yang--Baxter equation in its braid form,
$\RR$ playing the role of spectral set. We emphasize that $\cS$
depends in general
on $\chi_1$ and $\chi_2$ separately, which allows
\cite{Liguori:de} to treat systems with broken Lorentz (Galilean)
invariance ---
an expected feature when defects are present.

Recapitulating, with any solution $\cS$ of (\ref{unit1}), (\ref{qyb}) we
associate an associative algebra $\alg$, whose generators
$\{a_\alpha (\chi ),\, a^{\ast \alpha } (\chi ),\, r_\alpha^\beta
(\chi), \, t_\alpha^\beta (\chi)\}$
satisfy the constraints (\ref{aa})--(\ref{ta*}). The bulk exchange
relations (\ref{aa})--(\ref{aa*})
are similar to those of the ZF algebra, but for the
presence of the defect generators in the right-hand side of (\ref{aa*}).
The exchange properties of the latter are described by eqs.
(\ref{rr})--(\ref{tr}).
Equation (\ref{rr}) looks similar to
the boundary Yang--Baxter equation \cite{Cherednik:vs}, the difference
being that
in general the elements $\{r_\alpha^\beta (\chi ) \}$ do not commute
and, consequently, their position in (\ref{rr}) is essential.
Notice that $\{r_\alpha^\beta (\chi), \, t_\alpha^\beta (\chi)\}$
close a subalgebra of $\alg$, which generalizes the Sklyanin algebra
\cite{Sklyanin:yz} for the pure reflection case ($t_\alpha^\beta (\chi) = 0$).
The mixed relations (\ref{ar})--(\ref{ta*}) complete the list,
fixing the exchange properties between bulk and defect generators.

Two particular cases of $\alg$ were previously investigated.
Setting $t_\alpha^\beta (\chi) = 0$ one gets the boundary algebra
introduced in \cite{Liguori:1998xr} for handling integrable
systems on the half-line $\RR_+$. For
$r_\alpha^\beta (\chi) = t_\alpha^\beta (\chi) = 0$ one obtains instead
the ZF algebra, which applies to the same systems, but on the whole line
$\RR$. In this respect $\alg$ emerges as a unifying algebraic structure for
quantum field inverse scattering in 1+1 dimensions, which
works also in the presence of impurities. This expectation is widely
confirmed by the results reported in sect. 4.

In this paper we focus on RT algebras.
A RT algebra is a $\alg$-algebra whose defect generators satisfy in addition
\be
t_{\alpha_1}^\beta (\chi ) t^{\alpha_2}_\beta (\chi ) +
r_{\alpha_1}^\beta (\chi ) r^{\alpha_2}_\beta (-\chi ) =
\delta_{\alpha_1}^{\alpha_2} \, ,
\label{unit2}
\ee
\be
t_{\alpha_1}^\beta (\chi ) r^{\alpha_2}_\beta (\chi ) +
r_{\alpha_1}^\beta (\chi ) t^{\alpha_2}_\beta (-\chi ) = 0 \, .
\label{unit3}
\ee
A characteristic feature of any RT
algebra is a peculiar automorphism, which implements in algebraic
terms the physical
concepts of transmission and reflection and which is established a few
lines below.

{}For constructing the Fock representation of $\alg$, one needs
an involution. The most natural one is obtained by extending the mapping
\be
I \, : \,  a^{\ast \alpha }(\chi )  \mapsto a_\alpha (\chi )\, , \qquad \quad
I \, : \,  a_\alpha (\chi )  \mapsto a^{\ast \alpha }(\chi ) \, ,
\label{inv1}
\ee
\be
I \, : \,  r_\alpha^\beta  (\chi ) \mapsto r_\beta^\alpha (-\chi )\,
, \qquad \quad \, \, \,
I \, : \,  t_\alpha^\beta  (\chi ) \mapsto t_\beta^\alpha (\chi )\, ,
\label{inv2}
\ee
as an antilinear antihomomorphism on $\alg$. In fact, it is not
difficult to check that
$I$ leaves (\ref{aa})--(\ref{unit2}) invariant, provided that
\be
\cS_{12}^\dagger (\chi_1,\chi_2) = \cS_{12}(\chi_2,\chi_1)\, ,
\label{haS}
\ee
where the dagger stands for the Hermitian conjugation. The condition
(\ref{haS}),
known \cite{Zamolodchikov:xm} as Hermitian analyticity of $\cS$,
is assumed in what follows.

Let us finally consider the mapping
\be
\varrho \, : \, a_\alpha (\chi) \mapsto t_\alpha^\beta (\chi) a_\beta (\chi) +
r_\alpha^\beta (\chi) a_\beta (-\chi)\, ,
\label{aut1}
\ee
\be
\varrho \, : \, a^{\ast \alpha} (\chi) \mapsto a^{\ast \beta} (\chi)
t^\alpha_\beta (\chi) +
a^{\ast \beta} (-\chi) r^\alpha_\beta (-\chi) \, ,
\label{aut2}
\ee
\be
\varrho \, : \,  r_\alpha^\beta  (\chi ) \mapsto r^\beta_\alpha (\chi
)\, ,  \qquad \quad
\varrho \, : \,  t_\alpha^\beta  (\chi ) \mapsto t^\beta_\alpha (\chi )\, .
\label{aut3}
\ee
One can directly verify that $\varrho $ leaves
(\ref{aa})--(\ref{unit2}) invariant as well therefore
and extends  to an automorphism on $\alg$, considered as an
algebra with involution $I$.
We refer to $\varrho $ as reflection--transmission automorphism and
remark that because of
(\ref{unit2}) and (\ref{unit3}), $\varrho$ is idempotent. Equations
(\ref{aut1}) and (\ref{aut2})
are the algebraic counterparts of (\ref{c1}) and (\ref{c2}). They have
a simple physical interpretation:
each particle in the bulk is $\varrho$-equivalent to a superposition
of a transmitted and
reflected particle.

$\alg$ is an infinite algebra and from the above formal definition
it is not obvious at all that it has an operator
realization. Since such a realization is needed in the physical
applications, we will
construct in the next section an explicit representation of $\alg$ in
terms of (generally
unbounded) operators, which act in a Hilbert space.

\subsection{Fock representation}

We consider below representations of the RT algebra with involution
$\{\alg , I\}$ with the following structure:
\begin{description}

\item {(i)} the representation space is a complex Hilbert space $\Hil$ with
scalar product $(\cdot\, ,\, \cdot )$;

\item {(ii)} the generators
$\{a_\alpha (\chi ),\, a^{\ast \alpha } (\chi ),\, r_\alpha^\beta
(\chi ),\, t_\alpha^\beta (\chi )\}$
are operator-valued distributions with common and invariant dense
domain $\D\subset \Hil$, where eqs. (\ref{aa})--(\ref{unit3}) hold;

\item{(iii)} the involution $I$ is realized as a
conjugation with respect to $(\cdot\, ,\, \cdot )$.

\end{description}
A Fock representation is further specified by the condition:
\begin{description}

\item{(iv)} there exists a vacuum state $\Omega \in \D$,
which is annihilated by $a_\alpha (\chi)$. The vector $\Omega $ is
cyclic with respect to $\{a^{\ast \alpha } (\chi )\}$ and
$(\Omega\, ,\, \Omega ) = 1$.

\end{description}

There is a number of simple, but quite important consequences from
the assumptions
(i--iv). We start with
\medskip
 
\begin{prop}\label{prop1} The reflection--transmission
automorphism $\varrho $
is realized in any Fock representation by the identity operator.
\end{prop}

\noindent {\prf} We consider the matrix element
\be
(\varphi \, ,\, \{a_\alpha (\chi_1) -
\varrho [a_\alpha (\chi_1)]\} P[a^\ast]\Omega ) \, ,
\label{me1}
\ee
where $\varphi $ is an arbitrary state in $\D$ and $P$
is an arbitrary polynomial in $a^\ast$. Applying the identity
\be
\{a_\alpha (\chi_1) - \varrho [a_\alpha (\chi_1)]\}a^{\ast \beta}(\chi_2) =
a^{\ast \gamma}(\chi_2) \cS_{\alpha \gamma}^{\beta \delta}(\chi_1, \chi_2)
\{a_\delta (\chi_1) - \varrho [a_\delta (\chi_1)]\} \, ,
\label{id1}
\ee
which follows after some algebra from the exchange relations
(\ref{aa*}), (\ref{ra*}), (\ref{ta*}),
we can shift the curly bracket in (\ref{me1}) to the vacuum and
deduce from (iv) that
\be
(\varphi \, ,\, \{a_\alpha (\chi) -
\varrho [a_\alpha (\chi)]\} P[a^\ast]\Omega ) = 0\, .
\label{me2}
\ee
Taking the complex conjugate of (\ref{me2}), we obtain
\be
(P [a^*]\Omega\, ,\, \{a^{\ast \alpha} (\chi) -
\varrho [a^{\ast \alpha} (\chi)]\}\varphi ) = 0\, .
\label{me3}
\ee
{}Finally, using the cyclicity of $\Omega $, we conclude that
\be
a^{\ast \alpha} (\chi) = \varrho [a^{\ast \alpha} (\chi)]
= a^{\ast \beta} (\chi) t^\alpha_\beta (\chi) +
a^{\ast \beta} (-\chi) r^\alpha_\beta (-\chi)
\label{rti1}
\ee
holds on $\D$. Analogously, we derive
\be
a_\alpha (\chi) = \varrho [a_\alpha (\chi)]
= t_\alpha^\beta (\chi) a_\beta (\chi) +
r_\alpha^\beta (\chi) a_\beta (-\chi)\, .
\label{rti2}
\ee
Notice that the reflection--transmission identities (\ref{rti1}), (\ref{rti2})
generalize the $\delta$-impurity relations (\ref{c1}), (\ref{c2}).

\rightline{$\square$}

In what follows we are going to show that any RT algebra $\alg$
admits in general a whole
family $\rep$ of Fock representations, which can be parametrized by
means of the vacuum
expectation values
\be
\R_\alpha^\beta (\chi) = (\Omega\, ,\, r_\alpha^\beta (\chi) \Omega )
\, , \qquad
\T_\alpha^\beta (\chi) = (\Omega\, ,\, t_\alpha^\beta (\chi) \Omega ) \, ,
\label{vevTR}
\ee
called in what follows transmission and reflection matrices. Their
basic properties
are collected in

\begin{prop} \label{prop2}
      In each Fock representation of $\{\alg, I\}$:
\begin{description}
\item{(a)} $\T(\chi)$ and $\R(\chi)$ satisfy the Hermitian
analyticity conditions
\be
\R^\dagger (\chi) = \R(-\chi)\, ,
\label{haR}
\ee
\be
\T^\dagger (\chi) = \T(\chi)\, ;
\label{haT}
\ee

\item{(b)} the vacuum state $\Omega$ is unique (up to a phase
factor) and satisfies
\be
r_\alpha^\beta (\chi) \Omega = \R_\alpha^\beta (\chi) \Omega \, , \qquad \quad
t_\alpha^\beta (\chi) \Omega = \T_\alpha^\beta (\chi) \Omega \, .
\label{e1}
\ee

\item{(c)} $\T(\chi)$ and $\R(\chi)$ obey the
consistency relations
\bea
\cS_{12}(\chi_1 , \chi_2)\, \R_2(\chi_1)\,
\cS_{12}(\chi_2 , -\chi_1)\, \R_2(\chi_2) = \nonumber\\
\R_2(\chi_2)\, \cS_{12}(\chi_1 , -\chi_2)\, \R_2(\chi_1)\,
\cS_{12}(-\chi_2 , -\chi_1)\, ,
\label{rr1}
\eea
\bea
\cS_{12}(\chi_1 , \chi_2)\, \T_2(\chi_1)\,
\cS_{12}(\chi_2 , \chi_1)\, \T_2(\chi_2) = \nonumber\\
\T_2(\chi_2)\, \cS_{12}(\chi_1 , \chi_2)\, \T_2(\chi_1)\,
\cS_{12}(\chi_2 , \chi_1)\, ,
\label{tt1}
\eea
\bea
\cS_{12}(\chi_1 , \chi_2)\, \T_2(\chi_1)\,
\cS_{12}(\chi_2 , \chi_1)\, \R_2(\chi_2) = \nonumber\\
\R_2(\chi_2)\, \cS_{12}(\chi_1 , -\chi_2)\,
\T_2(\chi_1)\, \cS_{12} (-\chi_2 , \chi_1)\, ,
\label{tr1}
\eea
and unitarity conditions
\be
\T(\chi) \T(\chi) + \R(\chi) \R(-\chi) = \II\, ,
\label{unitTR1}
\ee
\be
\T(\chi) \R(\chi) + \R(\chi) \T(-\chi) = 0 \, .
\label{unitTR0}
\ee
\end{description}
\end{prop}
{\prf} The statement (a) is a direct consequence of
(\ref{inv1}) and point (iii) above.
Concerning (b), the argument implying the uniqueness of the vacuum is standard
(see e.g. \cite{RSII}). The identities in (\ref{e1}) can be deduced from
\be
([r_\alpha^\beta (\chi) - \R_\alpha^\beta (\chi)]\Omega\, ,\,
P[a^\ast]\Omega ) = 0
\label{me4}
\ee
and
\be
([t_\alpha^\beta (\chi) - \T_\alpha^\beta (\chi)]\Omega\, ,\,
P[a^\ast]\Omega ) = 0
\label{me5}
\ee
respectively, $P$ being an arbitrary polynomial.
{}For proving (\ref{me4}) and (\ref{me5}), one can shift by Hermitian
conjugation the polynomial to the first
factor and use afterwards the exchange relations (\ref{ar}) and (\ref{at})
and eq. (\ref{vevTR}).
Finally, (b) can be verified by taking the vacuum expectation values
of (\ref{rr})--(\ref{tr}) and
(\ref{unit2}), (\ref{unit3}) and using (\ref{e1}). This concludes the argument.

\rightline{$\square$}

We thus recover at the level of Fock representation the well-known
boundary Yang--Baxter equation (\ref{rr1}). A novel feature is the presence
of transmission (\ref{tt1}) and transmission--reflection (\ref{tr1})
Yang--Baxter equations.
Using (\ref{unit1}), eq. (\ref{tr1}) can be equivalently rewritten in the form
\bea
\cS_{12}(\chi_1 , \chi_2)\, \R_2(\chi_1)\,
\cS_{12}(\chi_2 , -\chi_1)\, \T_2(\chi_2) = \nonumber\\
\T_2(\chi_2)\, \cS_{12}(\chi_1 , \chi_2)\,
\R_2(\chi_1)\, \cS_{12} (\chi_2 , -\chi_1)\, .
\label{tr2}
\eea

Let us elaborate now a bit more on the relation between $\R$ and $\T$.
Because of (\ref{haR}), (\ref{haT}), $\T(\chi) \T(\chi)$ and $\R (\chi)
\R (-\chi)$
are non--negative Hermitian matrices which, according to (\ref{unitTR1}), are
simultaneously diagonalizable. The corresponding eigenvalues satisfy
\be
\lambda_i(\chi) + \mu_i(\chi) = 1\, ,\qquad \lambda_i(\chi) \geq 0\, ,
\quad \mu_i(\chi) \geq 0\, , \qquad i=1,...,N.
\label{eig}
\ee
Solving eq. (\ref{unitTR1}) for $\T$, one finds
\be
\T(\chi) =
\tau(\chi) \sqrt{\II - \R(\chi) \R(-\chi)} = \tau (\chi)\sum_{n=0}^\infty
\alpha_{n}\, [\R(\chi)\R(-\chi)]^n \, ,
\label{expT}
\ee
where $\tau (\chi)$ is some unknown function and the coefficients
$\alpha_{n}$ are determined by $\sqrt{1-x}=\sum_{n=0}^\infty \alpha_{n}x^n$.
The conditions (\ref{eig}) ensure that the series is convergent and
imposing (\ref{haT}), (\ref{unitTR1}) and (\ref{unitTR0}) on (\ref{expT}),
one obtains
\be
{\overline \tau}(\chi) = \tau (\chi)\, , \qquad \tau(\chi)^2 = 1\, ,
\qquad \tau(-\chi) = - \tau(\chi) \, .
\label{tau}
\ee
The series representation (\ref{expT}) of the matrix $\T$ allows to infer
the following remarkable property.

\begin{prop} \label{prop3} For any solution $\R$ of the boundary
Yang--Baxter equation {\rm(\ref{rr1})},
$\T$ defined by {\rm (\ref{expT})} satisfies {\rm (\ref{tt1}), (\ref{tr1})}.

\end{prop}
{\prf} The statement can be proven in two steps. The first one is to show that
the matrix $\T_{\R}(\chi) = \R(\chi) \R(-\chi)$ obeys (\ref{tt1}) and
(\ref{tr1}), which is done by repeated use of (\ref{rr1}). The second step
is based on the identities
\bea
\cS_{12}(\chi_1 , \chi_2)\, [\T_2(\chi_1)]^m\,
\cS_{12}(\chi_2 , \chi_1)\, [\T_2(\chi_2)]^n =
\nonumber\\ {[\T_2(\chi_2)]}^n\, \cS_{12}(\chi_1 , \chi_2)\, [\T_2(\chi_1)]^m\,
\cS_{12}(\chi_2 , \chi_1)\, ,
\label{Tid1}
\eea
\bea
\cS_{12}(\chi_1 , \chi_2)\, [\T_2(\chi_1)]^n\,
\cS_{12}(\chi_2 , \chi_1)\, \R_2(\chi_2) =
\nonumber\\
\R_2(\chi_2)\, \cS_{12}(\chi_1 , -\chi_2)\,
[\T_2(\chi_1)]^n\, \cS_{12} (-\chi_2 , \chi_1)\, ,
\label{Tid2}
\eea
which hold for any integers $m, n \geq 1$ and are the consequence
of a recurrent application of (\ref{tt1}), (\ref{tr1}).

\rightline{$\square$}

\noindent It is worth mentioning that the above argument makes no use
of the values of
coefficients $\alpha_n$ in (\ref{expT}) and the conclusion of 
proposition 3.3 remains
valid for any convergent series in powers of $\R(\chi)\R(-\chi)$.

Summarizing, we have shown that the transmission and
trans\-mission--reflection Yang--Baxter equations (\ref{tt1}) and (\ref{tr1})
are a consequence of
Hermitian analyticity (\ref{haR}), (\ref{haT}), unitarity 
(\ref{unit1}), (\ref{unitTR1}),
(\ref{unitTR0}) and
the boundary Yang--Baxter equation (\ref{rr1}).

We turn now to the Fock representations of $\alg$.
Our goal will be to demonstrate that each doublet
$\{\R,\, \T\}$, satisfying (\ref{haR}), (\ref{haT}),
(\ref{rr1}), (\ref{unitTR1}) and (\ref{unitTR0}), fully
determines a Fock representation $\frep$ of $\alg$. For this purpose
we shall construct
$\frep$ explicitly, extending the projection operator technique
developed in \cite{Liguori:1998xr, Liguori:1993pp, Liguori:1993uu} for the
ZF and boundary algebras, which are particular cases of $\alg$.
The first step is to introduce the $n$-particle subspace
$\Hil^{(n)}$ of $\frep$. For this purpose we consider
\be
\cL = \bigoplus_{\alpha} L^2(\RR) \, ,
\label{ds}
\ee
equipped with the standard scalar product
\be
(\varphi , \psi ) =
\int_{-\infty}^\infty d\chi\, \varphi^{\dagger \alpha } (\chi)
\psi_\alpha (\chi) =
\sum_{\alpha} \int_{-\infty}^\infty d\chi\, {\overline \varphi
}_\alpha (\chi) \psi_\alpha (\chi) \, .
\label{sp}
\ee
The $n$-particle space $\Hil^{(n)}$ we are looking for
is a subspace of the $n$-fold tensor power $\cL^{\otimes n}$,
characterized by a suitable projection operator $P^{(n)}$. In order to
construct $P^{(n)}$, we proceed as follows. Observing that any element
$\varphi\in \cL^{\otimes n}$
can be viewed as a column whose entries are
$\varphi_{\alpha_1 \cdots \alpha_n} (\chi_1,\dots,\chi_n)$, we
define the operators $\{\sigma^{(n)}_i, \, \tau^{(n)} \, \, :\, \,
i=1,..., n-1\}$
acting on $\cL^{\otimes n}$ according to:
\bea
&[\sigma^{(n)}_i\varphi ]_{\alpha_1 ... \alpha_n }
(\chi_1,...,\chi_i,\chi_{i+1},...,\chi_n ) = \qquad \qquad \nonumber \\
&\left [\cS_{i\,i+1}(\chi_i,\chi_{i+1})
\right ]_{\alpha_1 ... \alpha_n }^{\beta_1 ... \beta_n }
\varphi_{\beta_1 ... \beta_n }
(\chi_1,...,\chi_{i+1},\chi_i,...,\chi_n )\, , \qquad n\geq 2\, ,
\label{sigman}
\eea
\bea
\left[\tau^{(n)}\varphi \right ]_{\alpha_1 ... \alpha_n }
(\chi_1,...,\chi_n ) =
\T_{\alpha_n }^{\beta_n}(\chi_n)
\varphi_{\alpha_1 ... \alpha_{n-1}\beta_n}
(\chi_1,...,\chi_{n-1},\chi_n ) + \nonumber \\
\R_{\alpha_n }^{\beta_n}(\chi_n)
\varphi_{\alpha_1 ... \alpha_{n-1}\beta_n}
(\chi_1,...,\chi_{n-1},-\chi_n )\, , \qquad n\geq 1\, ,
\label{taun}
\eea
where
\be
\left [\cS_{ij}(\chi_i,\chi_j)
\right ]_{\alpha_1 ... \alpha_n }^{\beta_1 ... \beta_n } =
\delta_{\alpha_1}^{\beta_1}
\cdots { \widehat {\delta_{\alpha_i}^{\beta_i}}}
\cdots { \widehat {\delta_{\alpha_j}^{\beta_j}}}
\cdots \delta_{\alpha_n}^{\beta_n}
\, \cS_{\alpha_i \alpha_j}^{\beta_i \beta_j}(\chi_i,\chi_j) \, .
\label{def1}
\ee
{}In order to implement eqs. (\ref{sigman}), (\ref{taun}) on the whole
$\cL^{\otimes n}$,
we assume at this stage that the matrix elements
$\cS_{\alpha_1 \alpha_2 }^{\beta_1 \beta_2 }(\chi_1,\chi_2)$,
$\T_\alpha^\beta (\chi)$ and $\R_\alpha^\beta (\chi)$ are bounded functions.
Now, one can prove

\begin{prop}\label{prop4} Let $\W_n$
be the Weyl group associated with the root systems of
the classical Lie algebra $B_n$ and let
$\{\sigma_i,\, \tau\, \, :\, \, i=1,..., n-1\}$ be the generators of $\W_n$.
The mapping
\be
\phi_n\, :\, \sigma_i \mapsto \sigma_i^{(n)}\, , \qquad
\phi_n\, :\, \tau \mapsto \tau^{(n)}\, ,
\label{Wnrep}
\ee
defines a representation of $\W_n$ in $\cL^{\otimes n}$. Moreover,
\be
P^{(n)} \equiv {1\over 2^n n!}\,
\sum_{\nu \in \W_n} \, \phi_n(\nu )
\label{Pn}
\ee
is an orthogonal projection operator in $\cL^{\otimes n}$.
\end{prop}

\noindent{\prf} One has by construction
\bea
&\sigma_i^{(n)}\, \sigma_j^{(n)} =
\sigma_j^{(n)}\, \sigma_i^{(n)}\, , \qquad \quad |i-j|\geq 2 \, , \\
&\sigma_i^{(n)}\, \tau =\tau \, \sigma_i^{(n)} \, , \qquad \quad
1\leq i<n-2 \, .
\eea
Using the Yang--Baxter equations
(\ref{qyb}), (\ref{rr1})--(\ref{tr1}), (\ref{tr2}), one
shows that
\bea
\sigma_i^{(n)}\, \sigma_{i+1}^{(n)}\, \sigma_i^{(n)}  & =
\sigma_{i+1}^{(n)}\, \sigma_i^{(n)}\, \sigma_{i+1}^{(n)} \, ,  \\
\sigma_{n-1}^{(n)} \, \tau \, \sigma_{n-1}^{(n)} \, \tau &=
\tau \, \sigma_{n-1}^{(n)} \, \tau \, \sigma_{n-1}^{(n)} \, .
\eea
The unitarity conditions (\ref{unit1}) and (\ref{unitTR1}), 
(\ref{unitTR0}) imply
\be
[\sigma_i^{(n)}]^2 = \tau^2 = {\bf 1} \, .
\ee
Consequently, $\phi_n$ is a representation of $\W_n$ in $\cL^{\otimes n}$ and
$P^{(n)}$ is a projection operator. Finally, from Hermitian analyticity
(\ref{haS}), (\ref{haR}), (\ref{haT}), one infers that $\{\sigma_i^{(n)},\,
\tau^{(n)}\, \, :\, \, i=1,..., n-1\}$
are Hermitian operators. Therefore, $P^{(n)}$ is orthogonal.

\rightline{$\square$}

We have at this stage enough background to construct the Fock
representation
$\frep$. The $n$-particle space is defined by
\be
\Hil^{(0)} = \CC\, , \quad \qquad
\Hil^{(n)} = P^{(n)}\, \cL^{\otimes n}\, ,\quad  n\geq 1\, ,
\label{nps}
\ee
the total Fock space being
\be
\Hil = \bigoplus_{n=0}^\infty \Hil^{(n)} \, .
\label{totFs}
\ee
The finite particle space $\D$ is the (complex) linear space of sequences
\linebreak[4]
$\varphi = \left ( \varphi^{(0)}, \varphi^{(1)},...,\varphi^{(n)},...\right )$
with $\varphi^{(n)}\in \Hil^{(n)}$ and $\varphi^{(n)}=0$ for
$n$ large enough. $\D$ is dense in $\frep$.
The vacuum state is $\Omega = (1,0,...,0,...)$ and belong to $\D$.
The smeared bulk operators $\{a(f) , a^\ast (f) \, :\, f\in \cL \}$
act on $\D$ as follows:
\be
a(f) \Omega = 0\, ,
\label{aomega}
\ee
\be
\left [a(f) \varphi \right ]_{\alpha_1\cdots
\alpha_n}^{(n)}(\chi_1,...,\chi_n) =
\sqrt{n+1} \int_{-\infty}^\infty d\chi\, f^{\dagger \alpha_0 } (\chi)
\varphi_{\alpha_0 \alpha_1 \cdots \alpha_n}^{(n+1)} (\chi, \chi_1,...,\chi_n)
\, ,
\label{af}
\ee
\be
\left [a^\ast(f) \varphi \right ]_{\alpha_1\cdots
\alpha_n}^{(n)}(\chi_1,...,\chi_n)
= \sqrt {n} \left [P^{(n)} f\otimes \varphi^{(n-1)}
\right ]_{\alpha_1\cdots \alpha_n}(\chi_1,...,\chi_n) \, .
\label{a*f}
\ee
In general, $a(f) $ and $a^\ast(f) $ are unbounded operators on $\D$. For any
$\varphi^{(n)}\in \Hil^{(n)}$ one has however the estimate
\be
\parallel a^\natural (f) \varphi^{(n)} \parallel \, \leq \, \sqrt{n}
\parallel f
\parallel \parallel \varphi^{(n)} \parallel\, ,
\label{est1}
\ee
where $a^\natural (f)$ stands for $a(f)$ or $a^\ast (f)$.
Therefore $a(f)$ and $a^\ast (f)$ are bounded on each $\Hil^{(n)}$.

We now turn to the defect generators, defining $t_\alpha^\beta(\chi)$
and $r_\alpha^\beta(\chi)$ as the following multiplicative operators on $\D$:
\bea
&\!\!\!\!\left [r_\alpha^\beta(\chi) \varphi\right ]_{\gamma_1 ...
\gamma_n}^{(n)}(\chi_1,...,\chi_n) =
[\cS_{01}(\chi,\chi_1) \, \cS_{12}(\chi,\chi_2)\cdots \cS_{(n-1)\,
n}(\chi,\chi_n)\, \R_n(\chi)\cdot
\nonumber \\
&\cdot \cS_{(n-1)\, n}(\chi_n,-\chi)\cdots \cS_{12}(\chi_2,-\chi)
\, \cS_{01}(\chi_1,-\chi)]_{\alpha\gamma_1 ...
\gamma_n}^{\beta\delta_1 ... \delta_n}
\varphi^{(n)}_{\delta_1 ... \delta_n}(\chi_1,...,\chi_n) \, , \nonumber\\
\label{r}
\eea
\bea
&\left [t_\alpha^\beta(\chi) \varphi\right ]_{\gamma_1 ...
\gamma_n}^{(n)}(\chi_1,...,\chi_n) =
[\cS_{01}(\chi,\chi_1) \, \cS_{12}(\chi,\chi_2)\cdots \cS_{(n-1)\,
n}(\chi,\chi_n)\T_n(\chi)\cdot
\nonumber \\
&\cdot \cS_{(n-1)\, n}(\chi_n,\chi)\cdots \cS_{12}(\chi_2,\chi)
\, \cS_{01}(\chi_1,\chi)]_{\alpha\gamma_1 ...
\gamma_n}^{\beta\delta_1 ... \delta_n}
\varphi^{(n)}_{\delta_1 ... \delta_n}(\chi_1,...,\chi_n) \, ,
\quad\qquad
\label{t}
\eea
combined with (\ref{e1}). As expected, the defect operators
preserve the bulk particle number.

{}For deriving the commutation properties of
(\ref{af}), (\ref{a*f}), (\ref{r}), (\ref{t}) on $\D$, it
is convenient to introduce the operator-valued distributions
$a_\alpha (\chi)$ and $a^{\ast \alpha }(\chi)$ defined by
\be
a(f) = \int_{-\infty}^\infty d\chi\, f^{\dagger \alpha
}(\chi)a_\alpha (\chi) \, ,
\quad \quad
a^\ast (f)
= \int_{-\infty}^\infty d\chi\, f_\alpha (\chi)a^{\ast \alpha }(\chi) \, .
\label{ovd}
\ee
A straightforward computation allows to prove now

\begin{prop}\label{prop5} The operator-valued distributions
$\{a_\alpha (\chi),\, a^{\ast \alpha }(\chi) \}$ and\break
$\{r_\alpha^\beta(\chi), \, t_\alpha^\beta(\chi)\}$
satisfy the exchange relations {\rm (\ref{aa})--(\ref{ta*})} and the
constraints
{\rm (\ref{unit2})} and {\rm (\ref{unit3})} on $\D$. The involution $I$ is
realized as Hermitian conjugation with respect to the scalar product
{\rm (\ref{sp})}.
\end{prop}

This result completes the construction of the Fock representation $\frep$,
which is the basic tool in the physical applications discussed in
this paper. Fixing $\alg$,
$\frep$ is indeed fully determined by $\{\R, \T\}$ satisfying
eqs. (\ref{haR}), (\ref{haT}), (\ref{rr1})--(\ref{unitTR0}). Besides
some concrete examples,
little is known in general about the solution set of the latter.
There exist, however, one particular case
of physical importance, which is described in
\medskip

\begin{prop}\label{prop6} Suppose that
$\R$ obeys
\be
\cS_{12}(\chi_1,\chi_2)\R_2(\chi_1) = \R_1(\chi_1)\cS_{12}(-\chi_1,\chi_2)\, .
\label{cSR}
\ee
Then $\R$ satisfies the boundary Yang-Baxter equation {\rm (\ref{rr1})}.
Moreover, $\T$ obeys
\be
\cS_{12}(\chi_1,\chi_2)\T_2(\chi_1) = \T_1(\chi_1)\cS_{12}(\chi_1,\chi_2) \, ,
\label{cST}
\ee
and
\be
r_\alpha^\beta (\chi) \varphi = \R_\alpha^\beta (\chi) \varphi \, ,
\qquad \quad
t_\alpha^\beta (\chi) \varphi = \T_\alpha^\beta (\chi) \varphi \, ,
\label{e2}
\ee
hold for all $\varphi \in \D$.
\end{prop}
 
\noindent{\prf} Eq. (\ref{rr1}) follows directly from
(\ref{unit1}) and (\ref{cSR}). The identity (\ref{cST}) is a consequence of
(\ref{cSR}) and the series representation (\ref{expT}). Equations
(\ref{e2}) follow from
(\ref{t}), (\ref{r}) and (\ref{cSR}), (\ref{cST}). Note also that
(\ref{c1}), (\ref{c2}) are
recovered from (\ref{rti1}), (\ref{rti2}) and (\ref{e2}).

\rightline{$\square$}

The condition (\ref{cSR}) sort of linearizes eqs. (\ref{rr1})--(\ref{tr1})
and defines a special subset of representations ${\widetilde
\F}(\alg) \subset \rep$,
whose defect operators are proportional to the identity in $\Hil$.
All Fock representations of the algebras $\C_B$ and $\C_F$,
introduced above in the context of $\delta$-impurities, belong to
this subset because (\ref{cSR}) is identically satisfied for the
bosonic and fermionic exchange factors
\be
\cS_{\alpha_1 \alpha_2}^{\beta_1 \beta_2}(\chi_1, \chi_2) =
\pm\, \delta_{\alpha_1}^{\beta_2} \delta_{\alpha_2}^{\beta_1} \, .
\label{BF}
\ee
Let us observe in this respect that $\T$ and $\R$, given by
(\ref{TR1}), (\ref{TR2}), obey
Hermitian analyticity (\ref{haR}, \ref{haT}) and unitarity
(\ref{unitTR1}, \ref{unitTR0}).

\sect{Factorized scattering with impurities}

We develop in this section a general approach to factorized
scattering in (1+1)-dimensional
integrable models with impurity.

\subsection{Kinematics}

Let $E$ and $p$ be the energy and momentum of any asymptotic bulk particle.
Usually $E$ and $p$ are not independent and obey some dispersion relation.
The latter can be implemented expressing both $E$ and $p$
in terms of one parameter $\chi \in \RR$, namely
\be
E = E(\chi)\, , \qquad p = p(\chi)\, .
\label{gen}
\ee
It is instructive to keep in mind the following two examples:
\begin{itemize}
\item {} Relativistic dispersion relation
\be
E(\chi) = m \cosh (\chi)\, , \qquad p(\chi) = m \sinh (\chi)\, ,
\label{rel}
\ee
where $m>0$ is the mass and $\chi $ the rapidity.

\item {} Non-relativistic dispersion relation
\be
E(\chi) = \frac{m\chi^2}{2} + U\, , \qquad p(\chi) = m\chi \, ,
\label{nrel}
\ee
$\chi$ being the velocity and $U$ some constant.
\end{itemize}
Note that both of these relations satisfy
\be
\epsilon (p) = \epsilon (\chi)\, ,
\label{sign}
\ee
$\epsilon$ being the sign function. We also observe
   that a Lorentz boost in (\ref{rel}) and
a Galilean transformation in (\ref{nrel}) are both realized by a translation
$\chi \mapsto \chi + \alpha$.

In what follows we adopt a dispersion relation (\ref{gen}), which satisfies
(\ref{sign}) but is otherwise generic and parametrizes
any asymptotic bulk particle by $\chi \in \RR$ and its isotopic index
$i=1,...,N$.
Assuming that the impurity, localized at $x=0$, has no
internal degrees of freedom, and taking into account
(\ref{sign}), the fundamental building blocks of factorized scattering are:
\begin{description}

\item {(i)} the two-body bulk scattering matrix
$S_{i_1i_2}^{j_1j_2}(\chi_1,\chi_2)$
defined on $\RR \times \RR$;

\item {(ii)} the right and left reflection matrices $R^{+j}_{\, \, \,
\, \, \, i}(\chi )$
and $R^{-j}_{\, \, \, \, \, \, i}(\chi )$,
defined on $\RR_+$ and $\RR_-$ respectively and describing the
reflection of a particle from the impurity;

\item{(iii)} the left and right transmission matrices $T^{+j}_{\, \,
\, \, \, \, i}(\chi )$
and $T^{-j}_{\, \, \, \, \, \, i}(\chi )$,
defined on $\RR_+$ and $\RR_-$ respectively and describing the
transmission of a particle by the impurity.

\end{description}
We emphasize that $S$ is allowed to depend on $\chi_1$ and $\chi_2$ separately
\cite{Liguori:de}, generalizing the previous attempts
\cite{Delfino:1994nr, Konik:1997gx, Castro-Alvaredo:2002fc}, where
$S$ is assumed
to depend on $\chi_1-\chi_2$ only. As already argued in
\cite{Mintchev:2002zd}, this last condition
is too restrictive and quite artificial in the presence of defects.
With the dispersion relation (\ref{rel}) for instance,
$S_{i_1i_2}^{j_1j_2}(\chi_1-\chi_2)$ turns out to be Lorentz-invariant.
But we know that Lorentz symmetry is generally broken by impurities.
Accordingly, we
allow $S$ to depend on $\chi_1$ and $\chi_2$ separately. This
leads to a natural generalization of the inverse
scattering  method, which avoids the no-go theorem of
\cite{Delfino:1994nr, Castro-Alvaredo:2002fc} and
describes a large set of integrable systems, not covered there.

One should also keep in mind that our transmission and reflection matrices
are not defined on the whole $\RR$, but only for values of $\chi$
in the relative physical kinematic domains specified in (ii) and (iii).
This information must be sufficient for reconstructing
the total scattering operator $\bS$ and we demonstrate below that this is
indeed the case.

The data $\{S, R^\pm, T^\pm \}$ are subject to a number of constraints,
ensuring physical unitarity of the scattering operator $\bS$ and
factorization of the transition amplitudes. Let us first concentrate on
unitarity. Since integrability implies particle number conservation,
the restriction $\bS^{(1)}$ of $\bS$ to the one-particle subspace is
a well-defined
operator. One has (see also eq. (\ref{one-particle}) below)
\be
\bS^{(1)}(\chi) =\left(\begin{array}{cc}R^+(\chi)&T^+(\chi)\\
T^-(-\chi)&R^-(-\chi)\end{array}\right)\, ,\qquad \chi >0\, .
\label{S1}
\ee
Equation (\ref{S1}) reflects an essential difference with respect to any
Lorentz-invariant
theory, where $\bS^{(1)} = \II$ is mandatory \cite{RSIII}. Unitarity
\be
\bS^{(1)}(\chi) [\bS^{(1)}]^\dagger(\chi) =
[\bS^{(1)}]^\dagger(\chi) \bS^{(1)}(\chi) = \II
\label{1unit}
\ee
implies
\bea
&R^\pm(\pm \chi) [R^\pm]^\dagger (\pm \chi) +
T^\pm(\pm \chi) [T^\pm]^\dagger (\pm \chi) = \II \, ,
\label{1u} \\
&[R^\pm]^\dagger (\pm \chi) R^\pm(\pm \chi) +
[T^\mp]^\dagger (\mp \chi) T^\mp(\mp \chi) = \II \, ,
\label{2u} \\
&R^\pm(\pm \chi) [T^\mp]^\dagger (\mp \chi) +
T^\pm(\pm \chi) [R^\mp]^\dagger (\mp \chi) = 0 \, ,
\label{3u} \\
&[R^\pm]^\dagger (\pm \chi) T^\pm(\pm \chi) +
[T^\mp]^\dagger (\mp \chi) R^\mp(\mp \chi) = 0 \, ,
\label{4u}
\eea
where $\chi>0$.
We stress that (\ref{1u})--(\ref{4u}) are necessary and sufficient conditions:
any violation of (\ref{1u})--(\ref{4u}) breaks down
the unitarity of $\bS^{(1)}$ and, consequently, of $\bS$.
It is worth mentioning that in our previous paper \cite{Mintchev:2002zd}
\bea
&R^\pm(\pm \chi) R^\mp (\mp \chi) +
T^\pm(\pm \chi) T^\pm (\pm \chi) = \II \, ,
\label{1old} \\
&R^\pm (\pm \chi) T^\mp(\mp \chi) +
T^\pm (\pm \chi) R^\pm(\pm \chi) = 0 \, ,
\label{2old} \\
&[R^\pm ]^\dagger (\pm \chi) = R^\mp (\mp \chi) \, , \qquad
[T^\pm ]^\dagger (\pm \chi) = T^\pm (\pm \chi)\,
\label{3old}
\eea
were imposed instead of (\ref{1u})--(\ref{4u}). The conditions
(\ref{1old})--(\ref{3old}) are
stronger than (\ref{1u})--(\ref{4u}) and provide some technical
advantage \cite{Mintchev:2002zd}
in dealing with the factorization constraints obtained below. One can
easily see however that
$\delta$-type defects (see e.g. eq. (\ref{TR1}))
violate\renewcommand{\thefootnote}{*}\footnote{Correspondence with O.
A. Castro-Alvaredo and
A. Fring on this point is kindly acknowledged.}
(\ref{1old})--(\ref{3old}). For this reason we avoid
the use of (\ref{1old})--(\ref{3old}) in the present paper, keeping
(\ref{1u})--(\ref{4u}) which are respected
by the $\delta$-impurities described in sect. 2. We conclude the
issue recalling that
bulk scattering unitarity is controlled by
\be
S_{12}(\chi_1,\chi_2) S_{12}(\chi_2,\chi_1) = \II\, , \qquad
S_{12}^\dagger (\chi_1,\chi_2) = S_{12}(\chi_2,\chi_1) \, .
\label{Su}
\ee
 
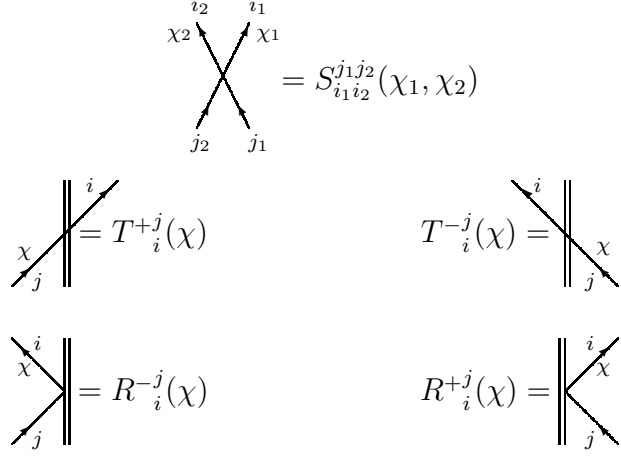
\begin{figure}[h]
\setlength{\unitlength}{0.7mm}
\begin{picture}(60,100)(-20,10)
%
%
\qbezier(65,80)(60,90)(55,100)
\qbezier(55,80)(60,90)(65,100)
\put(56.5,82.5){\vector(1,2){1}}
\put(64,97.5){\vector(1,2){1}}
\put(64,82.5){\vector(-1,2){1}}
\put(56.5,97.5){\vector(-1,2){1}}
\put(80,73){\makebox(20,20)[t]{$=S_{i_1i_2}^{j_1j_2}(\chi_1,\chi_2)$}}
\put(57,85){\makebox(20,20)[t]{${}_{i_1}$}}
\put(46,85){\makebox(20,20)[t]{${}_{i_2}$}}
\put(57,59){\makebox(20,20)[t]{${}_{j_1}$}}
\put(46,59){\makebox(20,20)[t]{${}_{j_2}$}}
\put(59,79){\makebox(20,20)[t]{${}_{\chi_1}$}}
\put(42,79){\makebox(20,20)[t]{${}_{\chi_2}$}}
%
%
\put (30,50){\line(0,0){20}}
\put (31,50){\line(0,0){20}}
\qbezier(20,50)(30,60)(40,70)
\put(22.5,52.5){\vector(1,1){1}}
\put(37.5,67.5){\vector(1,1){1}}
\put(35,44){\makebox(20,20)[t]{$=\Tp_i^j(\chi)$}}
\put(15,32.5){\makebox(20,20)[t]{${}_j$}}
\put(25,50.5){\makebox(20,20)[t]{${}_i$}}
\put(12.5,37.5){\makebox(20,20)[t]{${}_\chi$}}
%
%
\put (125,50){\line(0,0){20}}
\put (126,50){\line(0,0){20}}
\qbezier(135,50)(125,60)(115,70)
\put(132.5,52.5){\vector(-1,1){1}}
\put(118,67.5){\vector(-1,1){1}}
\put(100,44){\makebox(20,20)[t]{$\Tm_i^j(\chi)=$}}
\put(120,32.5){\makebox(20,20)[t]{${}_j$}}
\put(110,50.5){\makebox(20,20)[t]{${}_i$}}
\put(122.5,38.5){\makebox(20,20)[t]{${}_\chi$}}
%
%
%
\put (30,20){\line(0,0){20}}
\put (31,20){\line(0,0){20}}
\qbezier(20,20)(25,25)(30,30)
\qbezier(20,40)(25,35)(30,30)
\put(22.5,22.5){\vector(1,1){1}}
\put(22.5,37.5){\vector(-1,1){1}}
\put(35,14){\makebox(20,20)[t]{$=\Rm_i^j(\chi)$}}
\put(15,2.5){\makebox(20,20)[t]{${}_j$}}
\put(15,20.5){\makebox(20,20)[t]{${}_i$}}
\put(12.5,15.5){\makebox(20,20)[t]{${}_\chi$}}
%
%
\put (124,20){\line(0,0){20}}
\put (125,20){\line(0,0){20}}
\qbezier(135,20)(130,25)(125,30)
\qbezier(135,40)(130,35)(125,30)
\put(132.5,22.5){\vector(-1,1){1}}
\put(132.5,37.5){\vector(1,1){1}}
\put(100,14){\makebox(20,20)[t]{$\Rp_i^j(\chi)=$}}
\put(120,2.5){\makebox(20,20)[t]{${}_j$}}
\put(120,20.5){\makebox(20,20)[t]{${}_i$}}
\put(122.5,15.5){\makebox(20,20)[t]{${}_\chi$}}
\end{picture}
\caption{The two-body processes.}
\end{figure}

{}For analysing the constraints following from factorization, it is
convenient to
display the data $\{S, R^\pm, T^\pm \}$ graphically. This is done in
Fig. 1, where the time is flowing along the vertical direction, single lines
denote the particle world lines and the double line is the impurity.
Requiring factorization of all possible three-body processes
leads to a series of relations among $S$, $\Tpm$ and $\Rpm$.
As is well-known \cite{Zamolodchikov:xm}, the scattering of three
particles implies
the quantum Yang--Baxter equation
\be
S_{12}(\chi_1,\chi_2) S_{23}(\chi_1,\chi_3) S_{12} (\chi_2,\chi_3)
= S_{23}(\chi_2,\chi_3) S_{12}(\chi_1,\chi_3) S_{23}(\chi_1,\chi_2)  \, ,
\label{qyb1}
\ee
whose graphic representation is familiar and is omitted for conciseness.

The consistency conditions stemming from the scattering of two particles
between
themselves and the impurity, can be organized in three groups.

(a) Pure reflection:
\bea
&S_{12}(\chi_1, \chi_2)R^+_2(\chi_1)S_{12}(\chi_2 ,
-\chi_1)R^+_2(\chi_2) = \nonumber \\
&R^+_2(\chi_2)S_{12}(\chi_1, -\chi_2)R^+_2(\chi_1)S_{12}(-\chi_2, -\chi_1) \, ,
\label{SRSR+} \\
&S_{12}(\chi_1, \chi_2)R^-_1(\chi_2)S_{12}(-\chi_2 ,
\chi_1)R^-_1(\chi_1) = \nonumber \\
&R^-_1(\chi_1)S_{12}(-\chi_1, \chi_2)R^-_1(\chi_2)S_{12}(-\chi_2, -\chi_1) \, .
\label{SRSR-}
\eea
Equations (\ref{SRSR+}) and (\ref{SRSR-}) concern the reflection on $\hlp$ and
$\hlm$ respectively. Using the rules in Fig. 1 and moving back in time, one
gets the graphic representation of (\ref{SRSR+}) shown in Fig. 2.


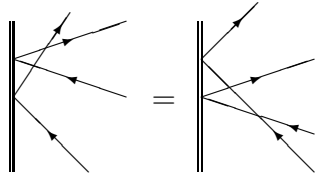
\begin{figure}[h]
\setlength{\unitlength}{.5mm}
\begin{picture}(50,40)(-60,0)
\put(35,0){\line(0,1){40}}
\put(34,0){\line(0,1){40}}
\put(65,20){\line(-3,1){30}}
\put(50,25){\vector(-3,1){1}}
\put(35,30){\line(3,1){30}}
\put(50,35){\vector(3,1){1}}
\put(55,0){\line(-1,1){20}}
\put(45,10){\vector(-1,1){1}}
%
\put(35,20){\line(2,3){15}}
\put(47,38){\vector(2,3){1}}
\put(65,5){\makebox(20,15)[t]{=}}
%
\put(85,0){\line(0,1){40}}
\put(84,0){\line(0,1){40}}
%
\put(115,10){\line(-3,1){30}}
\put(109,12){\vector(-3,1){1}}
\put(85,20){\line(3,1){30}}
\put(100,25){\vector(3,1){1}}
%
\put(115,0){\line(-1,1){30}}
\put(105,10){\vector(-1,1){1}}
\put(85,30){\line(1,1){15}}
\put(95,40){\vector(1,1){1}}
\end{picture}
\caption{Pure reflection.}
\end{figure}

\noindent The picture associated to (\ref{SRSR-}) is obtained from
Fig. 2 by reflection with respect
to the impurity world line.

(b) Pure transmission:
\bea
T^+_1(\chi_1)S_{12}(\chi_1, \chi_2)T^-_1(\chi_2) =
T^-_2(\chi_2)S_{12}(\chi_1, \chi_2)T^+_2(\chi_1) \, ,
\label{TST} \\
S_{12}(\chi_1, \chi_2)T^-_1(\chi_2)T^-_2(\chi_1) =
T^-_1(\chi_1)T^-_2(\chi_2)S_{12}(\chi_1, \chi_2)\, ,
\label{STT-} \\
S_{12}(\chi_1, \chi_2)T^+_1(\chi_2)T^+_2(\chi_1) =
T^+_1(\chi_1)T^+_2(\chi_2)S_{12}(\chi_1, \chi_2)\, .
\label{STT+}
\eea
Equations (\ref{TST}) and (\ref{STT-}) are represented in Fig. 3(a) and
Fig. 3(b) respectively.

\bigskip

\begin{figure}[h]
\setlength{\unitlength}{.5mm}
\begin{picture}(40,40)(-25,-10)
\put(150,0){\line(0,1){40}}
\put(149,0){\line(0,1){40}}
\put(165,12){\line(-3,1){40}}
\put(156,15){\vector(-3,1){0.1}}
\put(160,0){\line(-1,1){30}}
\put(155,5){\vector(-1,1){0.1}}
\put(165,5){\makebox(20,15)[t]{=}}
\put(165,-10){\makebox(20,15)[b]{(b)}}
%
\put(200,0){\line(0,1){40}}
\put(199,0){\line(0,1){40}}
%
\put(225,10){\line(-3,1){35}}
\put(219,12){\vector(-3,1){0.1}}
%
\put(225,0){\line(-1,1){35}}
\put(220,5){\vector(-1,1){0.1}}
%
%
\put(25,0){\line(0,1){40}}
\put(24,0){\line(0,1){40}}
\put(5,0){\line(1,1){30}}
\put(10,5){\vector(1,1){0.1}}
\put(35,0){\line(-1,1){30}}
\put(30,5){\vector(-1,1){0.1}}
\put(40,5){\makebox(20,15)[t]{=}}
\put(40,-10){\makebox(20,15)[b]{(a)}}
%
\put(75,0){\line(0,1){40}}
\put(74,0){\line(0,1){40}}
%
\put(65,0){\line(1,1){30}}
\put(70,5){\vector(1,1){0.1}}
%
\put(95,0){\line(-1,1){30}}
\put(90,5){\vector(-1,1){0.1}}
\end{picture}
\caption{Pure transmission.}
\end{figure}
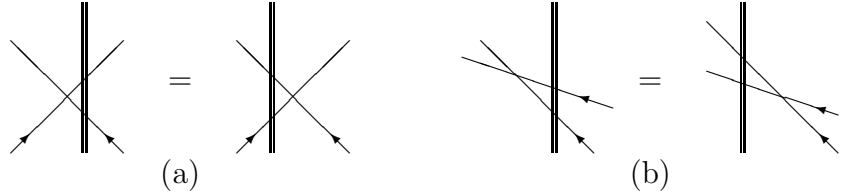

\noindent As before, the picture corresponding to eq. (\ref{STT+}) is
obtained from Fig. 3(b) by reflection.

(c) Mixed relations:
\bea
R^+_1(\chi_1)T^-_2(\chi_2) =
T^-_2(\chi_2)S_{12}(\chi_1, \chi_2)R^+_2(\chi_1)S_{12}(\chi_2, -\chi_1) \, ,
\label{TSRS+} \\
T^+_1(\chi_1)R^-_2(\chi_2) =
T^+_1(\chi_1)S_{12}(\chi_1, \chi_2)R^-_1(\chi_2)S_{12}(-\chi_2, \chi_1) \, ,
\label{TSRS-} \\
R^+_1(\chi_1)T^+_2(\chi_2) =
S_{12}(\chi_1, \chi_2)R^+_2(\chi_1)S_{12}(\chi_2, -\chi_1)T^+_2(\chi_2) \, ,
\label{SRST+} \\
T^-_1(\chi_1)R^-_2(\chi_2) =
S_{12}(\chi_1, \chi_2)R^-_1(\chi_2)S_{12}(-\chi_2, \chi_1)T^-_1(\chi_1)\, ,
\label{SRST-} \\
R^+_1(\chi_1)T^-_2(\chi_2)S_{12}(-\chi_1, \chi_2) =
T^-_2(\chi_2)S_{12}(\chi_1, \chi_2)R^+_2(\chi_1) \, ,
\label{TSR+} \\
T^+_1(\chi_1)R^-_2(\chi_2)S_{12}(\chi_1, -\chi_2) =
T^+_1(\chi_1)S_{12}(\chi_1, \chi_2)R^-_1(\chi_2) \, ,
\label{TSR-} \\
R^+_2(\chi_1)S_{12}(\chi_2, -\chi_1)T^+_2(\chi_2) =
S_{12}(\chi_2, \chi_1)R^+_1(\chi_1)T^+_2(\chi_2) \, ,
\label{RST+} \\
R^-_1(\chi_2)S_{12}(-\chi_2, \chi_1)T^-_1(\chi_1) =
S_{12}(\chi_2, \chi_1)T^-_1(\chi_1)R^-_2(\chi_2) \, .
\label{RST-}
\eea
Equations (\ref{TSRS+}) and (\ref{TSR+}) are shown in Fig. 4(a) and 4(b)
respectively,
whereas eqs. (\ref{SRST+}) and (\ref{RST+}) are drawn in Fig. 4(c) and 4(d).
The pictures related to the remaining four mixed equations are
obtained from Fig. 4 by
reflection, which completes the description of all three-body processes.

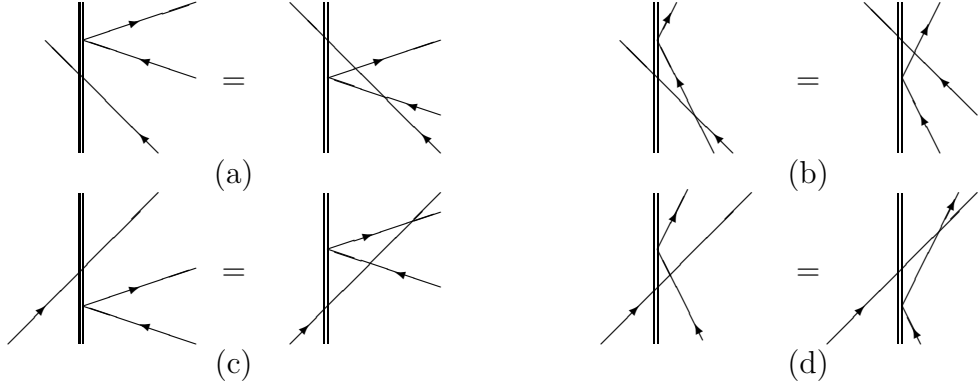
\begin{figure}[ht]
\setlength{\unitlength}{.5mm}
\begin{picture}(120,50)(10,-10)
\put(35,0){\line(0,1){40}}
\put(34,0){\line(0,1){40}}
\put(65,20){\line(-3,1){30}}
\put(50,25){\vector(-3,1){0.1}}
\put(35,30){\line(3,1){30}}
\put(50,35){\vector(3,1){0.1}}
\put(55,0){\line(-1,1){30}}
\put(50,5){\vector(-1,1){0.1}}
\put(65,5){\makebox(20,15)[t]{=}}
\put(65,-10){\makebox(20,15)[b]{(a)}}
%
\put(100,0){\line(0,1){40}}
\put(99,0){\line(0,1){40}}
%
\put(130,10){\line(-3,1){30}}
\put(121,13){\vector(-3,1){0.1}}
\put(100,20){\line(3,1){30}}
\put(115,25){\vector(3,1){0.1}}
%
\put(130,0){\line(-1,1){40}}
\put(125,5){\vector(-1,1){0.1}}
\end{picture}
%
%
\begin{picture}(120,50)(-20,-10)
\put(35,0){\line(0,1){40}}
\put(34,0){\line(0,1){40}}
\put(50,0){\line(-1,2){15}}
\put(40,20){\vector(-1,2){0.1}}
\put(35,30){\line(1,2){5}}
\put(38.5,37){\vector(1,2){0.1}}
\put(55,0){\line(-1,1){30}}
\put(50,5){\vector(-1,1){0.1}}
\put(65,5){\makebox(20,15)[t]{=}}
\put(65,-10){\makebox(20,15)[b]{(b)}}
%
\put(100,0){\line(0,1){40}}
\put(99,0){\line(0,1){40}}
%
\put(110,0){\line(-1,2){10}}
\put(105,10){\vector(-1,2){0.1}}
\put(100,20){\line(1,2){10}}
\put(107,34){\vector(1,2){0.1}}
%
\put(120,10){\line(-1,1){30}}
\put(110,20){\vector(-1,1){0.1}}
\end{picture}


\begin{picture}(140,50)(10,-10)
\put(35,0){\line(0,1){40}}
\put(34,0){\line(0,1){40}}
%
\put(15,0){\line(1,1){40}}
\put(25,10){\vector(1,1){0.1}}
\put(65,0){\line(-3,1){30}}
\put(50,5){\vector(-3,1){0.1}}
\put(35,10){\line(3,1){30}}
\put(50,15){\vector(3,1){0.1}}
\put(65,5){\makebox(20,15)[t]{=}}
\put(65,-10){\makebox(20,15)[b]{(c)}}
%
\put(100,0){\line(0,1){40}}
\put(99,0){\line(0,1){40}}
%
\put(130,15){\line(-3,1){30}}
\put(118,19){\vector(-3,1){0.1}}
\put(100,25){\line(3,1){30}}
\put(112,29){\vector(3,1){0.1}}
\put(90,0){\line(1,1){40}}
\put(95,5){\vector(1,1){0.1}}
\end{picture}
%
%
\begin{picture}(120,50)(0,-10)
\put(35,0){\line(0,1){40}}
\put(34,0){\line(0,1){40}}
\put(47,1){\line(-1,2){12}}
\put(45,5){\vector(-1,2){0.1}}
\put(35,25){\line(1,2){8}}
\put(40,35){\vector(1,2){0.1}}
\put(20,0){\line(1,1){40}}
\put(30,10){\vector(1,1){0.1}}
\put(65,5){\makebox(20,15)[t]{=}}
\put(65,-10){\makebox(20,15)[b]{(d)}}
%
\put(100,0){\line(0,1){40}}
\put(99,0){\line(0,1){40}}
%
\put(80,0){\line(1,1){40}}
\put(90,10){\vector(1,1){0.1}}
\put(105,0){\line(-1,2){5}}
\put(103,4){\vector(-1,2){0.1}}
\put(100,10){\line(1,2){15}}
\put(114,38){\vector(1,2){0.1}}
\end{picture}

\caption{Mixed relations.}
\end{figure}

Summarizing, the scattering data $\{S, R^\pm, T^\pm \}$ are required to satisfy
two sets of conditions: unitarity constraints
(\ref{1u})--(\ref{4u}), (\ref{Su}) and
factorization constraints (\ref{qyb1})--(\ref{RST-}). The general
solution of this
long list of matrix equations is currently unknown. In order to
simplify the problem,
we consider invertible $T^\pm$. From eqs. (\ref{TSR+}) and
(\ref{RST-}) one then infers that
\be
S_{12}(\chi_1,\chi_2)R^\pm_2(\chi_1) =
R^\pm_1(\chi_1)S_{12}(-\chi_1,\chi_2)\, ,
\label{SR}
\ee
which implies the validity of all
(\ref{SRSR+}), (\ref{SRSR-}), (\ref{TSRS+})--(\ref{RST-}).
Therefore, assuming that
$T^\pm$ are invertible, one is left with
eqs. (\ref{1u})--(\ref{4u}), (\ref{Su}), (\ref{qyb1}),
(\ref{TST})--(\ref{STT+}) and
(\ref{SR}), which simplifies a little the problem of deriving
explicit solutions
$\{S, R^\pm, T^\pm \}$. At this stage, it might be useful to give some
examples,
starting with the $gl(N)$-invariant $S$-matrix
\be
S_{12}(\chi_1,\chi_2)=\frac{1}{s(\chi_1) - s(\chi_2) +i\, g}
\left\{\left [s(\chi_1) - s(\chi_2)\right ]\, P_{{12}}+ig \,
\II\otimes \II \right\} \, ,
\label{Sgln}
\ee
where $P_{12}$ is the standard flip operator, $g\in \RR$, and
$s(\chi)$ is any real valued
{\sl even} function. For $R^\pm$ and $T^\pm$ one derives
\bea
&R^\pm (\chi ) = [\cos p(\pm \chi )]\exp [iq^\pm(\pm \chi)]\, \II\, ,
\label{Rgln}\\
&T^\pm (\chi ) = \pm [\sin p(\pm \chi )]\exp [iq^\pm(\pm \chi)]\, \II \, ,
\label{Tgln}
\eea
$p(\chi )$ and $q^\pm(\chi )$ being arbitrary real valued functions on $\RR_+$.
In this example both reflection and transmission preserve the
isotopic type and
all isotopic types have the same reflection and transmission amplitude.

A more complicated example is provided by the Toda type $S$-matrix
\be
S_{i_1i_2}^{j_1j_2}(\chi_1,\chi_2) = \exp\left
[is_{i_1i_2}(\chi_1,\chi_2)\right ]\,
\delta_{i_1}^{j_2} \delta_{i_2}^{j_1} \, ,
\label{Stoda}
\ee
where $s_{i_1i_2}(\chi_1,\chi_2)$ are real valued functions obeying
\be
s_{i_1i_2}(\chi_1,\chi_2) = - s_{i_2i_1}(\chi_2,\chi_1)\, , \qquad
s_{i_1i_2}(\chi_1,\chi_2) = s_{i_1i_2}(\chi_1,-\chi_2)\, .
\label{todas}
\ee
When $s_{i_1i_2}(\chi_1,\chi_2)$ satisfy (\ref{todas}) but are
otherwise generic, one finds
\be
[R^\pm]_i^j(\chi ) = [\cos p_i(\pm \chi )]\exp [iq_i^\pm (\pm
\chi)]\, \delta_i^j\, ,
\label{todaR}
\end{equation}
\be
[T^\pm ]_i^j(\chi ) = \pm [\sin p_i(\pm \chi )]\exp [iq_i^\pm (\pm
\chi)]\, \delta_i^j\, ,
\label{todaT}
\end{equation}
where $p_i(\chi )$ and $q_i^\pm(\chi )$ are real valued functions on $\RR_+$.
Also here the impurity interaction preserves the isotopic type, but
the individual
reflection and transmission amplitudes may be different. Finally, if
some of the
entries $s_{i_1i_2}(\chi_1,\chi_2)$ coincide, non-diagonal elements
in $R^\pm$ and $T^\pm$
are allowed \cite{Mintchev:2002zd} and the isotopic type is not preserved.

\subsection{Scattering operator and transition amplitudes}

We have so far described in great detail the main features of the physical data
$\{S, R^\pm, T^\pm \}$ for factorized scattering with impurity. The
next step is to identify
the RT algebra $\alg$ and its Fock representation $\frep$ producing
the total scattering operator $\bS$ and the transition amplitudes,
corresponding to
$\{S, R^\pm, T^\pm \}$. For this purpose we set
\be
\cS (\chi_1 , \chi_2) =
\left (\matrix {
S(\chi_1,\chi_2) & 0 & 0 & 0 \cr
0 & 0 & S(\chi_1,\chi_2) & 0 \cr
0 & S(\chi_1,\chi_2) & 0 & 0 \cr
0 & 0 & 0 & S(\chi_1,\chi_2)\cr} \right ) \, ,
\label{cS}
\ee
\be
\R(\chi ) =
\theta(\chi)\left(\begin{array}{cc}R^+(\chi)
&0\\ 0&{[R^-]}^\dagger(-\chi)\end{array}\right)
+
\theta(-\chi)\left(\begin{array}{cc}{[R^+]}^\dagger(-\chi)
&0\\ 0&R^-(\chi)\end{array}\right) ,
\label{cR}
\ee
\bea
\T(\chi ) =
\theta(\chi)\left(\begin{array}{cc}0&T^+(\chi) \\
{[T^+]}^\dagger(\chi)&0\end{array}\right) +
\theta(-\chi)\left(\begin{array}{cc}0&{[T^-]}^\dagger(\chi)\\
T^-(\chi)&0\end{array}\right) ,\qquad
\label{cT}
\eea
$\cS$, $\R$ and $\T$ defined above are admissible because of

\begin{prop}\label{prop7} The constraints {\rm (\ref{1u})--(\ref{4u}),
(\ref{Su}), (\ref{qyb1})}
and {\rm (\ref{SR})} on the data $\{S, R^\pm, T^\pm \}$ imply the validity of
{\rm (\ref{unit1}), (\ref{qyb}), (\ref{haS}), (\ref{haR}), (\ref{haT}),
(\ref{unitTR1}), (\ref{unitTR0}), (\ref{cSR})} for $\{\cS, \R, \T \}$.
\end{prop}

\noindent{\prf} We first observe that the condition of
Hermitian analyticity
(\ref{haR}), (\ref{haT}) for $\T$ and $\R$ is satisfied by construction.
The remaining conditions can be checked by direct computation.

\rightline{$\square$}

Thus, eq. (\ref{cS}) determines the algebra $\alg$, whereas
(\ref{cR}), (\ref{cT}) fix the
representation $\frep $ in terms of $\{R^\pm, T^\pm \}$.
We stress that eq. (\ref{SR}) implies (\ref{cSR}). Therefore,
according to proposition \ref{prop6}, $\frep \in {\widetilde
\F}(\alg)$. In other words,
the factorization conditions derived in sect. 4.1 select representations from
the subclass ${\widetilde \F}(\alg) \subset  \rep$.

The asymptotic states in $\frep $ are defined in complete analogy
with the $\delta$-impurity
case, discussed in sect. 2. The presence of internal degrees of freedom can
be dealt with in a straightforward way. In-states are created from the vacuum
by $\{a^{\ast (-,i)}(g)\, :\, \supp g \subset \RR_+\}$ and
$\{a^{\ast (+,i)}(g)\, :\, \supp g \subset \RR_-\}$.
The out-states are generated instead by
$\{a^{\ast (-,j)}(h)\, :\, \supp h \subset \RR_-\}$ and
$\{a^{\ast (+,j)}(h)\, :\, \supp h \subset \RR_+\}$.
By means of (\ref{a*a*}), one can also order the
creation operators according to the values of the spectral parameter,
using the relation $\prec$ introduced in sect. 2. We thus
define the ``in"-coming states by
\be
|g_1,i_1;...;g_m,i_m\rangle^{\rm in} =
a^{\ast (\xi_1, i_1)}(g_1)\cdots a^{\ast (\xi_m, i_m)}(g_m) \Omega \, ,
\label{in1}
\ee
where
\be
g_1 \prec \cdots \prec g_m \, , \qquad
\xi_i =
\left\{ \begin{array}{cc}
+ \, , & \quad \mbox{$\supp g_i \subset \RR_-$}\, ,
\\ -\, , & \quad \mbox{$\supp g_i \subset \RR_+$}\, .
\end{array} \right.
\label{non1}
\ee
The ``out"-going states are given by
\be
{}^{\rm out}\langle h_1,j_1;...;h_n,j_n| =
a^{\ast (\eta_1, j_1)}(h_1)\cdots a^{\ast (\eta_n, j_n)}(h_n) \Omega \, ,
\label{out1}
\ee
with
\be
h_n \prec \cdots \prec h_1 \, , \qquad
\eta_j =
\left\{ \begin{array}{cc}
+ \, , & \quad \mbox{$\supp h_j \subset \RR_+$}\, ,
\\ -\, , & \quad \mbox{$\supp h_j \subset \RR_-$}\, .
\end{array} \right.
\label{non2}
\ee
The asymptotic spaces $\Hil^{\rm in}$ and $\Hil^{\rm out}$ are
generated by finite
linear combinations of vectors of the type (\ref{in1}) and
(\ref{out1}) respectively.
Each of these spaces is dense in $\Hil$. The total
scattering operator ${\bf S}\, :\, \Hil^{\rm out}\to \Hil^{\rm in}$
is defined by
\be
{\bf S}\, :\, a^{\ast (\eta_1, j_1)}(h_1)\cdots a^{\ast (\eta_n,
j_n)}(h_n) \Omega
\longmapsto a^{\ast ({\widetilde \eta}_1, j_1)}({\widetilde h}_1)\cdots
a^{\ast ({\widetilde \eta}_n, j_n)}({\widetilde h}_n) \Omega \, ,
\label{Stot}
\ee
where
\be
{\widetilde h}_k (\chi ) = h_k( -\chi)\, , \qquad {\widetilde \eta}_k
= - \eta_k \, .
\label{tilde}
\ee
Using the non-overlapping conditions (\ref{non1}), (\ref{non2}), it is
not difficult to check that
\be
\left({\bf S}\Psi^{\rm out}\, ,\, {\bf S}\Phi^{\rm out}\right) =
\left(\Psi^{\rm out}\, ,\, \Phi^{\rm out}\right) \, , \qquad
\forall \Psi^{\rm out},\,  \Phi^{\rm out} \in \Hil^{\rm out} \, .
\label{unittot}
\ee
Generalizing the argument of \cite{Liguori:1998xr, Liguori:de},
we deduce from (\ref{Stot}), (\ref{unittot}) that ${\bf S}$ is unitary.

A generic scattering amplitude reads
\bea
&{}^{\rm out}\langle h_1,j_1;...;h_n,j_n
|g_1,i_1;...;g_m,i_m\rangle^{\rm in} = \nonumber \\
&\left (a^{\ast (j_1,\eta_1)}(h_1)\cdots a^{\ast (j_n,\eta_n)}(h_n) \Omega \, ,
a^{\ast (i_1,\xi_1)}(g_1)\cdots a^{\ast (i_m,\xi_m)}(g_m ) \Omega \right )
\label{trampl}
\eea
and can be computed by means of the exchange relation (\ref{aa*}) and
the identities (\ref{e1}).
The Fock structure implies that (\ref{trampl}) vanishes unless $m=n$,
showing the absence of particle production as expected from integrability.
The one-particle transition amplitudes can be deduced from
the correlation function
\bea
&\left (a^{\ast \beta}(\chi)\Omega \, , \,
a^{\ast \alpha}(\vph) \Omega \right ) =   \nonumber \\
&\left [\delta^{\alpha}_{\beta}+
{{\T}}^{\alpha}_{\beta}(\chi)\right ]\delta(\chi-\vph)+
{{\R}}^{\alpha}_{\beta}(\chi) \delta(\chi+\varphi) \, .
\label{two-point function}
\eea
One gets
\be
\!\!\!
{}^{\rm out}\langle h,j|g,i\rangle^{\rm in}  =
\left\{ \!\!\!\!\!\!\!\!
\begin{array}{cc}
\int_0^\infty d\chi \overline h(\chi) {T^+}^{i}_{j}(\chi)g(\chi),
& \quad \mbox{$\xi=-,\, \eta=+$}\, ,\\[1ex]
\qquad \int_0^\infty d\chi \overline h(-\chi) {T^-}^{i}_{j}(-\chi)g(-\chi),
& \quad \mbox{$\xi=+,\, \eta=-$}\, ,\\[1ex]
\int_0^\infty d\chi \overline h(\chi) {R^+}^{i}_{j}(\chi)g(-\chi),
& \quad \mbox{$\xi=+,\, \eta=+$}\, , \\[1ex]
\quad \int_0^\infty d\chi \overline h(-\chi) {R^-}^{i}_{j}(-\chi)g(\chi),
& \quad \mbox{$\xi=-,\, \eta=-$}\, ,
\end{array} \right.
\label{one-particle}
\end{equation}
which describe the particle--impurity interaction and precisely reproduce
the one-particle scattering matrix $\bS^{(1)}$ given by eq. (\ref{S1}).

The particle--particle interaction shows up
in the two-particle amplitudes, which can be derived from the correlator
\bea
\lefteqn{
\left (a^{\ast \beta_1}(\chi_1) a^{\ast \beta_2}(\chi_2 ) \Omega \, ,
\, a^{\ast \alpha_1}(\vph_1) a^{\ast \alpha_2}(\vph_2 ) \Omega \right
) \ =\   }\nonu
&&[\delta^\mu_{\beta_{2}}+{{\T}}^\mu_{\beta_{2}}(\chi_{2})]\,
{\cS}_{\beta_{1}\mu}^{\alpha_{1}\nu}(\chi_{1},\chi_{2})\,
[\delta^{\alpha_{2}}_{\nu}+{{\T}}^{\alpha_{2}}_{\nu}(\chi_{1})]\
\delta(\chi_{1}-\varphi_{2})\, \delta(\chi_{2}-\vph_{1})\nonu
&+&{{\R}}^\mu_{\beta_{2}}(\chi_{2})\,
{\cS}_{\beta_{1}\mu}^{\alpha_{1}\nu}(\chi_{1},-\chi_{2})\,
[\delta^{\alpha_{2}}_{\nu}+{{\T}}^{\alpha_{2}}_{\nu}(\chi_{1})]\
\delta(\chi_{1}-\varphi_{2})\, \delta(\chi_{2}+\vph_{1})\nonu
&+&[\delta^\mu_{\beta_{2}}+{{\T}}^\mu_{\beta_{2}}(\chi_{2})]\,
{{\cS}}_{\beta_{1}\mu}^{\alpha_{1}\nu}(\chi_{1},\chi_{2})\,
{{\R}}^{\alpha_{2}}_{\nu}(\chi_{1})\
\delta(\chi_{1}+\varphi_{2})\, \delta(\chi_{2}-\vph_{1})\nonu
&+&{{\R}}^\mu_{\beta_{2}}(\chi_{2})\,
{{\cS}}_{\beta_{1}\mu}^{\alpha_{1}\nu}(\chi_{1},-\chi_{2})\,
{{\R}}^{\alpha_{2}}_{\nu}(\chi_{1})\
\delta(\chi_{1}+\varphi_{2})\, \delta(\chi_{2}+\vph_{1})\nonu
&+&[\delta^{\alpha_1}_{\beta_1}+\T^{\alpha_1}_{\beta_1}(\chi_1)]\,
[\delta^{\alpha_2}_{\beta_2}+\T^{\alpha_2}_{\beta_2}(\chi_2)]\
\delta(\chi_1-\varphi_1)\, \delta(\chi_2-\vph_2)\nonu
&+&[\delta^{\alpha_1}_{\beta_1}+\T^{\alpha_1}_{\beta_1}(\chi_1)]\,
\R^{\alpha_2}_{\beta_2}(\chi_2)\
\delta(\chi_1-\varphi_1)\, \delta(\chi_2+\vph_2)\nonu
&+&\R^{\alpha_1}_{\beta_1}(\chi_1)\,
[\delta^{\alpha_2}_{\beta_2}+\T^{\alpha_2}_{\beta_2}(\chi_2)]\
\delta(\chi_1+\varphi_1)\, \delta(\chi_2-\vph_2)\nonu
&+&\R^{\alpha_1}_{\beta_1}(\chi_1)\,
\R^{\alpha_2}_{\beta_2}(\chi_2)\
\delta(\chi_1+\varphi_1)\, \delta(\chi_2+\vph_2)\, .
\label{4-point}
\eea
Take for instance the asymptotic states
\be
|g_1,i_1;g_2,i_2\rangle^{\rm in}\, , \qquad \xi_1 = -,\quad \xi_2 = + \, ,
\label{examplin}
\end{equation}
\be
{}^{\rm out}\langle h_1,j_1;h_2,j_2|\, , \qquad \eta_1 = +, \quad
\eta_2 = + \, .
\end{equation}
The corresponding transition amplitude receives contributions only from
the second and the third term in the
right-hand side of (\ref{4-point}). One finds
\bea
&{}^{\rm out}\langle h_1,j_1;h_2,j_2|g_1,i_1;g_2,i_2\rangle^{\rm in}
= \nonumber \\
&\int_0^\infty d\chi_1 d\chi_2 {\overline h_1}(\chi_1) {\overline h_2}(\chi_2)
[{{R^+}}^k_{j_{2}}(\chi_{2}){S}_{j_{1}k}^{i_{1}l}(\chi_{1},-\chi_{2})
{{T^+}}^{i_{2}}_{l}(\chi_{1})g_1(-\chi_2) g_2(\chi_1) + \nonumber \\
&{R^+}^{i_1}_{j_1}(\chi_1){T^+}^{i_2}_{j_2}(\chi_2) g_1(-\chi_1)
g_2(\chi_2)] \, .
\eea
The associated scattering processes are displayed in Fig. 4 (c,d).
All possible kinematic domains, respecting the
non-overlapping conditions (\ref{non1}), (\ref{non2}), give rise to
nine different two-particle transition
amplitudes, which are reported in \cite{Mintchev:2002zd}.

Summarizing, the physical scattering data $\{S, R^\pm, T^\pm \}$
determine both the
RT algebra $\alg$ and its Fock representation $\frep$ entering the
derivation of the
$\bf S$-matrix amplitudes. The asymptotic states are obtained by
acting with the particle creation
operators on the standard Fock vacuum $\Omega \in \frep$. It is worth
stressing that
our scheme makes no use of any auxiliary construction of a {\sl
boundary state} with
prescribed reflection and transmission properties.
This essential difference with respect to all previous approaches
of the subject \cite{Ghoshal:tm}--\cite{Castro-Alvaredo:2002dj}
represents a relevant theoretical
and technical advantage of the framework based on the RT algebra $\alg$.

We emphasize, in conclusion, that the above scattering theory is based
entirely on the
data $\{S, R^\pm, T^\pm \}$ for {\sl real} values of the spectral parameter
$\chi$. For this reason the results of this work are very general and
remain valid also after imposing
all physical conditions (such as crossing symmetry and certain
meromorphic structure)
on the continuation of $\{S, R^\pm, T^\pm \}$ to the complex $\chi$-plane.

\sect{Conclusions and perspectives}

We developed in this article a framework for dealing with factorized scattering
from reflecting and transmitting impurities in 1+1 dimensions.
Our starting point was the analysis of some exactly solvable models
with $\delta$-type
impurities, which led us directly to the main tool of our
approach, the RT algebra $\alg$.
The interaction of a particle with the impurity is implemented in
$\alg$ by the reflection and transmission generators
$\{r_\alpha^\beta (\chi)\}$
and $\{t_\alpha^\beta(\chi)\}$ respectively.
As already mentioned, setting $t_\alpha^\beta (\chi)
= 0$, one gets from $\alg$ another useful algebra $\balg$, which
describes \cite{Liguori:1998xr}
factorized  scattering from a purely reflecting boundary. In this
context $\balg$ applies also
to the construction of off-shell correlation functions
\cite{Gattobigio:1998hn, Gattobigio:1998si} and to the study of symmetries
\cite{Mintchev:2001aq}. Moreover, setting $r_\alpha^\beta (\chi) =
t_\alpha^\beta (\chi) = 0$,
one obtains the celebrated ZF algebra. Therefore, $\alg$
indeed represents a universal structure for dealing with integrable models
in 1+1 dimensions.

The Fock representations of $\alg$ also exhibit remarkable features.
The operators $\{r_\alpha^\beta (\chi),\, t_\alpha^\beta (\chi)\}$ condense in
the vacuum $\Omega \in \frep$. The relative condensates
$\{\R_\alpha^\beta (\chi),\, \T_\alpha^\beta(\chi)\}$ are
directly related to the physical reflection and  transmission amplitudes.
There is no need for special boundary or reflection--transmission states
in our scheme. The use of the standard Fock vacuum $\Omega \in \frep$ for
deriving the asymptotic states significantly simplifies the construction.

We established a complete set of factorization conditions for scattering
with impurities in 1+1 dimensions, showing that they
admit solutions with non-trivial bulk scattering if
the requirement of Lorentz (Galilean) invariance on the bulk $S$-matrix is
relaxed. This feature, which is not surprising in the presence of
defects, allows the no-go theorem of \cite{Castro-Alvaredo:2002fc}
to be avoided.

The concept of RT algebra, introduced in this paper, opens a variety
of new directions for further
research. On the mathematical side, the interplay between $\alg$,
$\balg$ and ZF algebras
deserves a more detailed analysis. A link between $\balg$ and the ZF
algebra has been
explored in \cite{Ragoucy:2001cf}. From the physical point of view,
$\alg$ appears to be the natural
candidate for replacing the ZF algebra in the form-factor program for
integrable models with
impurities. The construction of off-shell local fields in this context is
a challenging open problem. We strongly believe that besides to
integrable systems,
RT algebras apply also to (1+1)-dimensional conformal field theory
with permeable
walls, which partly transmit and partly reflect the incident waves.
Such theories
\cite{Bachas:2001vj, DeWolfe:2001pq} have obvious relevance to
critical phenomena and have recently acquired
some importance in the theory of strings and branes. Finally, for
applications to
impurity problems in condensed-matter physics, one needs finite temperature
representations of the RT algebras, which requires the construction
of Kubo--Martin--Schwinger
states over $\alg$. We are currently investigating some of these issues.


\end{document}